**Retail electricity costs and emissions incentives are misaligned for commercial and industrial power consumers**

Fletcher T. Chapin[a], Akshay K. Rao[a], Adhithyan Sakthivelu[a], Carson I. Tucker[b], Eres David[c], Casey S. Chen[d], Erin Musabandesu[a], Meagan S. Mauter[a,e,f,g,h,*]

[a]*Department of Civil and Environmental Engineering, Stanford University, 473 Via Ortega, Stanford, California 94305, United States*
[c]*Department of Mechanical Engineering, Stanford University, 440 Escondido Mall Building 530, Stanford, California 94305, United States*
[c]*Department of Earth and Environmental Engineering, Columbia University, 500 W 120th St, New York, New York 10027, United States*
[d]*Symbolic Systems Program, Stanford University, 389 Jane Stanford Way, Stanford, California 94305, United States*
[e]*Department of Environmental Social Sciences, Stanford University, 473 Via Ortega, Stanford, California 94305, United States*
[f]*Senior Fellow, Woods Institute for the Environment, Stanford University, 473 Via Ortega, Stanford, California 94305, United States*
[g]*Senior Fellow, Precourt Institute for Energy, Stanford University, 473 Via Ortega, Stanford, California 94305, United States*
[h]*Photon Science, SLAC National Accelerator Laboratory, 2575 Sand Hill Road, Menlo Park, California 94025, United States*

**Corresponding author: mauter@stanford.edu*

## Abstract

Electrification is contributing to substantial growth in U.S. commercial and industrial loads, but the cost and Scope 2 carbon emission implications of this load growth are opaque for both power consumers and utilities. This work describes a unique spatiotemporally resolved data set of U.S. electricity costs and emissions and applies time series approximation methods to quantify the alignment of electricity cost and emission incentives for large commercial and industrial consumers. We present a comprehensive spatiotemporal dataset of U.S. price-based demand response (i.e., tariff) and incentive-based demand response programs, enabling direct comparison to previously published marginal emission factor, average emission factor, and day-ahead market prices. We resolved the structural incompatibility and fragmentation of these datasets by developing time series approximations of discrete data and unifying geospatially heterogeneous datasets. Analysis of these datasets reveals significant spatial and temporal heterogeneity in cost and carbon emissions incentives for demand-side energy flexibility, underscoring the importance of site selection as a key factor influencing power costs and scope 2 emissions. Analysis also reveals broad misalignment of economic and emissions incentives under existing electricity tariff structures, meaning tariffs are incentivizing consumption of more carbon-intensive electricity, and highlighting potential barriers to electrification delivering carbon savings.

## Introduction

Electrification is shifting emissions and costs of energy consumption from Scope 1 (onsite primary energy consumption) to Scope 2 (grid-supplied electrical power). In a fully decarbonized and economically efficient electricity market, complete electrification eliminates $CO_2$ emissions and financially incentivizes power consumption profiles that avoid generation and transmission constraints (Sofia and Dvorkin, 2024). In reality, the significant variance in marginal emissions factors across grid regions and time-of-day obscure emissions reduction benefits of electrification (de Chalendar and Benson, 2019; Suri et al., 2024), while complex electricity tariff structures complicate the optimization of power consumption profiles (Chapin et al., 2024). Indeed, past work has highlighted cases from carbon intensive or over-constrained power networks where electrification increased $CO_2$ emissions (Jiang et al., 2025; Tamayao et al., 2015) and the price of power (Goteti et al., 2021; Lin et al., 2021).

Realizing best-case electrification scenarios during the grid transition depends on incentivizing large industrial and commercial loads to shift consumption away from carbon intensive and grid constrained (peak) times of day and grid nodes. The societal benefits of doing so are significant: the U.S. DOE identified opportunities for industrial energy flexibility to avoid $5-35 billion in U.S. grid modernization costs, reduce renewables curtailment by up to 60 TWh, and reduce carbon emissions by 44.4 million metric tons $CO_2$-eq (Sonali Razdan et al., 2025; Zhou and Mai, 2021). The benefits to large industrial and commercial load consumers who provide these energy services, however, have been much more challenging to estimate (Gorka et al., 2025). Doing so requires spatially and temporally resolved data on electricity wholesale prices, tariffs structures, incentive-based demand response programs, and Scope 2 average and marginal emissions factors that are not currently available in a unified format or location.

In the absence of a unified, machine-readable dataset, it has also been challenging to determine whether financial incentives for load shifting also incentivize Scope 2 carbon emissions reductions. Previous studies have focused on a single independent system operator (ISO), such as CAISO (Mayes and Sanders, 2022; Neutel et al., 2026; Wei et al., 2013) or PJM (Blumsack, 2009), or a single customer class, such as single family homes (Mayes and Sanders, 2022). Additionally, existing studies about the alignment of costs and emissions tend to focus on future grid scenarios for grid planning purposes (Borenstein and Kellogg, 2023; Evro et al., 2026; Kök et al., 2018; Wei et al., 2013), while our study focuses on quantifying existing cost and emissions alignment. This is a key difference, as grid projections include levers for grid planning, such as investment in new energy resources and retirement of aging technology. Our method allows for quantification of existing cost and emissions alignment without knowledge of future grid operating decisions.

Quantifying the alignment of electricity costs and carbon emissions would also help determine the implications of load growth, load siting, and load shaping. The U.S. is anticipating significant load growth from a combination of electrification and new demand (e.g., data centers), but policy makers lack national data on where and when that growth is substantively changing grid emissions. Industrial and commercial operators could also use spatiotemporally resolved electricity and emissions data for siting new loads to minimize cost and/or exploit energy arbitrage opportunities. Similarly, this data is critical for determining the cost-effectiveness of facility electrification, guiding the installation of energy flexibility upgrades at facilities, and calculating the energy performance characteristics of those upgrades (e.g., round-trip efficiency, power capacity, energy capacity) (Rao et al., 2024).

Finally, grid operators and public utility commissions need spatiotemporally resolved cost and carbon data for evaluating price-based (i.e., tariff) and incentive-based flexible program design. Unified cost and emissions data is critical to determining the effectiveness of legislative mandates to internalize environmental externalities (i.e., carbon emissions of generation) in rate setting. This data can also be used to determine the extent to which retail tariffs, demand response programs, and wholesale electricity markets are providing consistent incentives to large load consumers. Grid operators may also be interested in using unified data to explore whether price-based demand response (PBDR) or incentive-based demand response (IBDR) programs effectively align cost and emissions incentives and how baselining policies for IBDR programs influence that balance.

Existing cost and emissions datasets have facilitated answers to some of those questions, but data incompatibility and fragmentation limit the scope and relevance of insights into others. Average emissions factors are now reported by the U.S. Energy Information Administration, with imports and exports between balancing authorities recognized, using a method developed by de Chalendar et al. (de Chalendar et al., 2019; de Chalendar and Benson, 2021). Marginal emissions factors can be computed using a method developed by Siler-Evans et al. (Siler-Evans et al., 2012). Similarly, national datasets on wholesale market costs and emissions are available from Grid Status (GridStatus.io, 2025) and WattTime (Callaway et al., 2018; WattTime, 2025). OpenEI provides tariffs collected from the United States Utility Rate Database (USURDB) (McLaren et al., 2017; OpenEI, 2025). Recent scholarly work has compiled the tariffs applicable to 100 specific large load customers, accompanied by cost calculation code (Chapin et al., 2024). Finally, data on demand response (DR) programs is available from the U.S. DOE (U.S. Department of Energy, 2024a).

The incompatibility of these datasets limits analysis of the spatiotemporal dynamics of cost, carbon emissions, and their alignment. Each of these datasets are represented by diverse data structures (e.g., discrete, continuous, or combinations thereof) that operate on different time scales (e.g., defined period maximums, interval averages, marginal signals). For example, electricity tariffs have energy charges based on hourly consumption and demand charges based on maximum consumption over the billing period. Time-of-use (TOU) tariffs have multiple discrete (possibly overlapping) periods. IBDR datasets are even more complex; each program has unique baseline and payment procedures that are frequently nonlinear. This data incompatibility makes it difficult to quantify alignment between incentive structures. For example, economic subsidies for hydrogen production in the form of tax credits, which are intended to achieve net-zero goals in difficult to decarbonize sectors, have the potential to increase emissions (Ricks et al., 2023).

Fragmentation of existing data further impedes comprehensive national analyses of the spatiotemporal dynamics of power cost and carbon intensity. Even when data would be otherwise compatible (i.e., they can be represented by the same data structure), it can be difficult to resolve at the same geographic or temporal timescale. For example, wholesale electricity prices may be available at the node-level, while average emissions factors are reported by balancing authority. Or wholesale electricity prices may be reported every 15 minutes, while emissions factors are reported hourly. Besides geographic and temporal fragmentation, some datasets have different customer classifications (e.g., residential, commercial, or industrial), while other datasets aggregate customers and report a single value. In conclusion, evaluating economic and emissions incentive structures depends on integrating data that span diverse temporal and spatial scales.

As a result of both incompatibility and fragmentation, past analysis has struggled to deliver generalizable insight into where, when, and to what degree cost and emissions incentive structures are aligned. As a workaround, case study analyses tend to interrogate alignment questions using a single wholesale market or tariff structure (Puschnigg et al., 2023), but are unable to quantitatively assess their representativeness vis-à-vis regional or national grids.

We propose a novel data reconciliation approach for unifying previously incompatible and fragmented electricity price and emissions datasets. We then provide a comprehensive national dataset of electricity costs (tariffs and IBDR programs) that is compatible with existing average and marginal emissions factor data. Finally, we apply this national database of electricity cost and emissions data to analyze the alignment of incentives using the reconciled data.

## Methods

### Overview

We collect data for Scope 2 emissions factors, wholesale and retail electricity prices, and incentive-based demand response (IBDR) programs (Figure 1). Marginal emissions factors (MEFs), average emission factors (AEFs), and wholesale (i.e., day-ahead market (DAM)) prices are aggregated from existing datasets (GridStatus.io, 2025; U.S. Energy Information Administration, 2023; U.S. Environmental Protection Agency, 2023). Retail electricity prices (i.e., tariffs) are converted from the USURDB (McLaren et al., 2017; OpenEI, 2025), validated for correctness, and merged with manually collected tariffs (Chapin et al., 2024). IBDR parameters are collected from the U.S. DOE (U.S. Department of Energy, 2024a), converted to our new, machine-readable data format, and published as the Incentive Demand Response Program Parameter (IDRoPP) dataset (David et al., 2024).

After data is collected, we reconcile incompatible and fragmented data using methods specific to each dataset. We then perform two separate analyses on the reconciled data: correlation and regime mapping. All data used throughout the paper is for the year 2023, but the data collection, reconciliation, and analysis methods are readily applied to other years using code maintained at

https://github.com/we3lab/electric-emissions-misalignment. Additionally, we recently released Electric Emissions Cost & Optimization (EECO), a Python package that enables using the data presented in this paper to calculate electrical costs and emissions from static data or within optimization models (Chapin et al., 2025b).

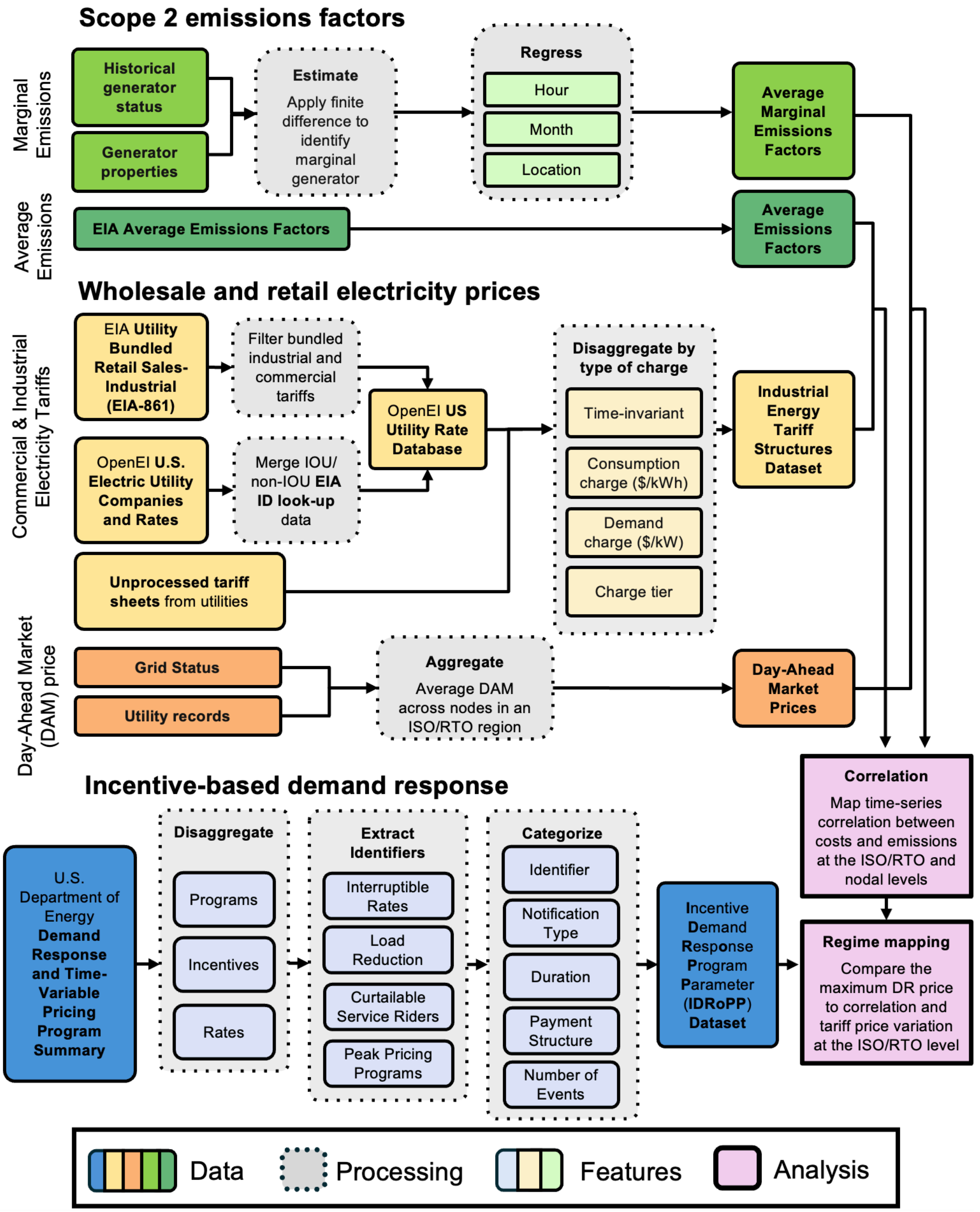

**Figure 1: Information flow for dataset curation.** Data covers marginal (light green) and average (dark green) emissions factors, retail (yellow) and wholesale (orange) electricity prices, and incentive-based demand response programs (blue). The leftmost box represents publicly available or purchasable datasets. Processing steps are highlighted by grey boxes. Features that are used to regress, average, or resample original datasets are highlighted in lighter colored solid boxes within the processing steps. Finally, pink boxes on the right describe the analysis conducted within this paper. The machine-readable data structures used in the IDRoPP and electricity tariff datasets are an original contribution of this work.

Data acquisition

*Scope 2 emissions factors*

Average emission factors (AEFs) represent the average emissions intensity of all generators across the grid region at a given moment in time, making them an appropriate point of reference for estimating the emissions intensity of pre-existing, inflexible loads. Marginal emission factors (MEFs) represent the emission intensity of the marginal generator at a given moment in time. In other words, MEFs account for the emissions impact of producing (or reducing) an additional kWh of electricity. Therefore, MEFs are most appropriate when decisions by a consumer will lead to a change in grid demand, such as energy flexible operation or the siting of new loads.

To provide insight for planning and operational contexts for US grid regions, we compute MEF data as a monthly/hourly average and use historical AEF time series from 2023. We calculate MEFs using the regression algorithm developed by Siler-Evans et al. (Siler-Evans et al., 2012). Their method identifies the marginal generator for a region based on historical generation data, and then computes the average marginal emission factor for each hour of the month (U.S. Environmental Protection Agency, 2023). Their algorithm can be applied to compute MEFs by eGRID region, ISO and regional transmission operators (RTO), or state. In this manuscript, we use hourly average MEFs from 2023 calculated for each ISO/RTO.

Historical AEFs are reported semiannually by the U.S. Energy Information Administration (U.S. Energy Information Administration, 2023) using a reconciliation method from de Chalendar et al. (de Chalendar et al., 2019; de Chalendar and Benson, 2021). We collect historical hourly AEF time series for the entirety of 2023 directly from the EIA. The EIA data is reported by balancing authority, which includes ISOs and RTOs.

*Wholesale and retail electricity prices*

Electricity markets can be broadly classified under wholesale and retail pricing categories. Wholesale prices represent the cost of electricity purchased directly from the ISO, without any distribution utility, including real-time and near real-time markets (e.g., DAM). We use the DAM price as the wholesale price throughout this study. We collect DAM data directly from Grid Status where available (GridStatus.io, 2025). Data for the Midcontinent Independent System Operator (MISO) and Southwest Power Pool (SPP) regions were incomplete in Grid Status for 2023, so we gathered the data directly from the MISO and SPP websites (Midcontinent Independent System Operator, 2023; Southwest Power Pool, 2023). We did not collect DAM data from outside the major ISOs because it would be illogical to average the geospatially disparate sources. For example, a composite average DAM across the southeastern and northwestern United States when the regions comprise of different electricity markets administered by different entities would not be meaningful. Unlike the other data in this study, we are unable to publish the DAM data since it is proprietary. We recommend that others obtain raw data directly from Grid Status (GridStatus.io, 2025), WattTime (WattTime, 2025), or another third-party service.

Retail prices are the prices paid by typical consumers, usually published in the form of electricity tariffs. The United States Utility Rate Database (USURDB) (OpenEI, 2025) aggregates >50,000 tariffs and is the starting point for our analysis. We filter the USURDB to 1,492 tariffs using the following criteria:

- Sector: Industrial or Commercial
- Service Type: Bundled or Delivery with Standard Offer[1]
- Start Date: Before January 1, 2023
- End Date: After January 1, 2023
- Minimum Peak Capacity (kW): ≤ 1 MW
- Minimum Peak Capacity (kW): ≥ 1 MW

We then cross-reference the EIA ID for each utility in "U.S. Electric Utility Companies and Rates: Look-up by Zipcode (2023)" to determine ZIP code (Huggins, 2024). For easier geolocation, we use `pgeocode` ("pgeocode v0.5.0," 2024) to map between ZIP codes and latitude and longitude coordinates using the GeoNames database (Wick, 2025). The GeoNames database is relatively accurate when accounting for the "snapping" of lat/long coordinates to a grid, which leads to an error of about 1.85 km (Ahlers, 2013). This is sufficiently accurate for our use case of mapping utility service areas, which are far wider than 1.85 km.

We convert the data structure of the remaining 1,492 tariffs from the USURDB into an extended version of our previously published data format (Chapin et al., 2024) modified to accommodate additional features. For example, we added an "assessed" column to denote whether demand charges are assessed monthly or daily, as daily demand charges are becoming more common in regions like CAISO.

Converting the data format was essential to fully describing all tariffs; in other words, the set of tariffs represented by our format is a superset of the tariffs that can be represented by OpenEI's data format. Converting the data format also facilitates human readability (by converting the tariff data of >50,000 entries from a single spreadsheet to individual CSV files) and compatibility with automated workflows (by precluding the need to load the entire dataset into memory at once). To facilitate this, we provide a `metadata.csv` file for users to find the relevant tariff ID

[1] There is also a "Delivery Only" version of the dataset denoted by the suffix "delivery_only" in the file names (as opposed to the suffix "bundled")

based on latitude/longitude, EIA ID, etc. Finally, converting the data format also enabled development of the Electric Emissions & Cost Optimizer (EECO) Python package for computing the electricity bill for any tariff in our data format, given a load profile (Chapin et al., 2025b). While we only use a small set of EECO's features in this paper, the converted tariffs can be used with EECO's more advanced features to model and optimize industrial operations.

The complete conversion code is available on GitHub at https://github.com/we3lab/industrial-electricity-tariffs/. For this analysis, we focus on bundled electricity tariffs, meaning that generation, transmission, and distribution charges are included in the price. In some service areas, however, delivery and generation are provided by separate utilities. For those regions, we also provide a delivery only version of the dataset. The dataset contains both versions denoted by the suffix "delivery_only" or "bundled" in the file names.

*Incentive-based demand response*

We use the U.S. DOE Federal Energy Management Program's Demand Response and Time-Variable Pricing website (U.S. Department of Energy, 2024a), which lists and describes energy management programs across the United States, to create the Incentive Demand Response Program Parameter (IDRoPP) dataset (David et al., 2024).

First, we define program parameters (a subset of which are shown in Table 1). These parameters characterize each program by its distinct features. We select these parameters based on practical experience with demand response simulation to include necessary attributes that fully define and model IBDR events (Sakthivelu et al., 2025).

**Table 1: Subset of Incentive Demand Response Program Parameter (IDRoPP) fields.** The parameters in this table come from a comprehensive list of all program parameters included in the IDRoPP Dataset (David et al., 2024). These parameters are included because they are the most important standard descriptors of these programs.

| Parameter | Description |
|---|---|
| Notification type (\`notif_type\`) | The DR event notification type (e.g., day ahead or the day of) |
| Baseline calculation method (\`base_method\`) | The method used by each program to calculate the baseline energy usage |
| Payment function (\`pay_function\`) | Form of compensation for shifts in energy load (or being available to shift energy loads) when called upon |
| Region (\`region\`) | The United States Department of Energy separates the states into the West Region, The Southeast and Midwest Region, and the Northeast Region |
| State (\`state\`) | State in which the program is eligible |
| Company (\`comp\`) | Which company or companies offer this program |

| Trigger (`trigger`) | Event trigger type |
|---|---|

Next, we populate the dataset with programs along with their parameters (if available) and a URL link to the program source/website. The database population consists of three substeps:

1. Disaggregate: The original dataset aggregates IBDR data into “programs”, “incentives”, and “rates”, so we disaggregate them into a single category for all IBDR programs.
2. Extract identifiers: gather the relevant identifiers from each program, such as interruptible rates, curtailable service riders, and peak pricing programs.
3. Categorize: re-label the extracted data based on the identifier so that it matches the IDRoPP format.

After populating the database, we cross-validate IDRoPP with the U.S. DOE Federal Energy Management Program's Demand Response and Time-Variable Pricing Programs Search (U.S. Department of Energy, 2024b). Users can also contribute their own data in the IDRoPP format for IBDR programs absent from the U.S. DOE database, subject to author review. Further instructions on contributing are available at https://github.com/we3lab/dr-programs.

Data reconciliation

*Scope 2 emissions factors*

The Scope 2 emission factors (i.e., AEFs and MEFs) are compatible with one another because both are represented in a timeseries format. However, the linear regression method from Siler-Evans et al. returns an hourly average (Siler-Evans et al., 2012), whereas the EIA reports a historical month-long time series (U.S. Energy Information Administration, 2023). In this study, we averaged the historical AEF by month and hour to reconcile the MEF and AEF data.

*Wholesale and retail electricity prices*

The raw data for wholesale and retail electricity prices is both incompatible and fragmented. For example, the tariff demand charges are priced based on maximum monthly consumption, while DAM prices are reported as a historical time series. To address this incompatibility, we convert the electricity tariffs to time series using the following method:

1. We select a flat 1 MW base load profile and use the EECO package to create a linear charge function for the load profile to calculate the total electricity cost (Chapin et al., 2025b).
2. Distribute the demand and energy charges evenly over the active time steps. By active time steps, we mean periods where the TOU charge is non-zero.
3. Sum the distributed charges within each time step.

While we use a 1 MW base load for this analysis, our method works for any desired load. The generalizability of our method allows one to convert any tariff to time series format given an assumed load, regardless of how the load was discretized (i.e., timestep size, load shape, etc.). Analysis of other load shapes from ELMAS (Bellinguer et al., 2023) can be found in the SI (*S.3.4. Alternative Load Shapes*).

After addressing data incompatibilities due to inherent data structure differences, we address data fragmentation issues. We average all prices by month and hour, yielding a representative 24-hour time series for each month to address temporal fragmentation. Geospatial fragmentation is slightly more complicated to address. First, we obtain ISO geographic information system (GIS) data (specifically the `Independent_System_Operators.shp`) from the Department of Homeland Security's Homeland Infrastructure Foundation-Level Data (HIFLD) (U.S. Department of Homeland Security, 2025). Second, we use that GIS data to match the latitude and longitude in the tariff dataset to ISO. If a tariff is in a location that does not belong to an ISO, the ISO label is set to "Other" and removed from analysis.

*Incentive-based demand response*

The IBDR data is not fully reconcilable with other electricity-related emissions and costs because of key distinctions in incentive structure. IBDR programs are optional bid-based programs largely available to commercial and industrial loads, whereas market-based price signals apply to all consumers and are cleared at a set time in advance (e.g., day-ahead, 1-hour, or 5-minute). Tariffs are renegotiated periodically (e.g., yearly) and are available to charge consumers at every minute until renegotiated.

IBDR program rules and the decision to initiate an event are dependent on the grid operator or program managers' decision to engage participants. For now, it is fundamentally different from the other price incentives because the incentive is not available to all participants most of the time. While it can be put in a timeseries, the largest factor in the value of IBDR participation is characterizing the uncertainty around the timing and duration of the IBDR event for a given participant. In summary, tariffs, wholesale prices, and Scope 2 emissions factors can all be computed for a flat 1 MW load, but IBDR requires additional operational simulation to stochastically model IBDR events.

<u>Data analysis</u>

The data analysis in this paper is a brief demonstration of the myriad applications that are unlocked by our data reconciliation approach. We investigate the peak-to-off-peak ratios within a single tariff charge using the aggregated tariff timeseries to understand trends in how tariffs incentivize load shifting. We also compute the Pearson correlation coefficients between cost and emissions signals to quantify how aligned these incentive signals are. When the Pearson correlation coefficient is positive, costs are high when emissions are high (and vice versa). It is not possible to compute the Pearson correlation coefficient on raw tariff and emissions data, however, because it requires two vectors of equal length. We use the `corrcoef` function from NumPy to compute the correlation coefficient after data reconciliation (Harris et al., 2020).

## Results

<u>Dataset characterization</u>

*Scope 2 emissions factors*

We draw upon existing datasets (U.S. Energy Information Administration, 2023; U.S. Environmental Protection Agency, 2023) and methods (de Chalendar et al., 2019; de Chalendar and Benson, 2021; Siler-Evans et al., 2012) to plot the monthly spread in average and marginal emissions factors on an hourly basis in 2023 (Figure 2). Taking CAISO as an example, the AEF varies from hour-to-hour between 240 and 346 kg $CO_2$-eq/MWh in the month of January (Figure 2a). This decreases to 136 to 259 kg $CO_2$-eq/MWh in the spring months when solar output peaks. Later in the summer (July-September) the solar intensity of the grid remains high, but increased temperatures also increase demand on the grid (e.g., for air conditioning) (Mayes et al., 2024).

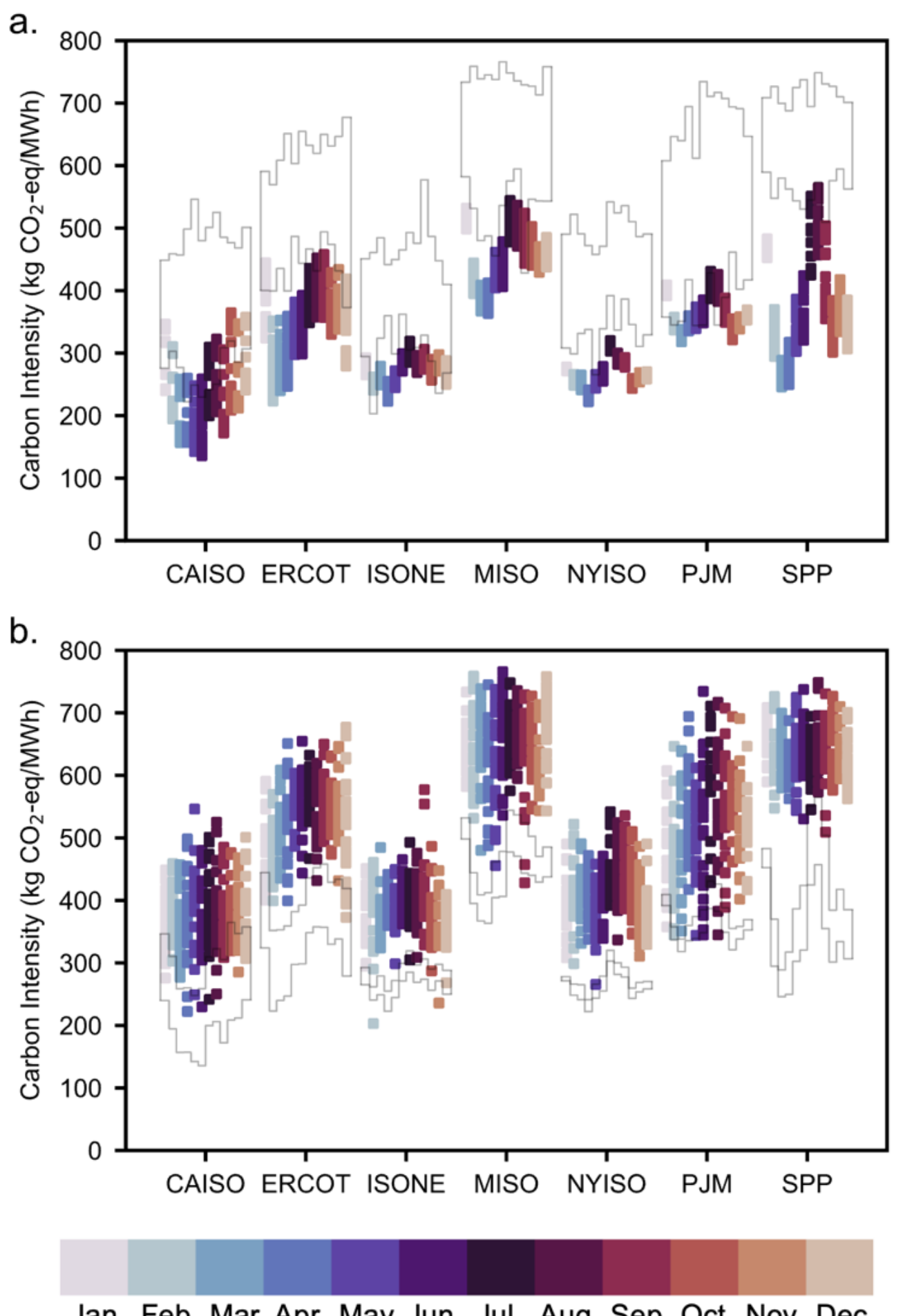

**Figure 2: Grid electrical emissions factors in 2023.** a). Average emissions defined as the average Scope 2 emissions per hour divided by the average power generated in that hour. b) Marginal emissions defined by the emissions associated with the generator that will supply the next unit of energy for a given hour. Points on each plot represent the hourly average intensities for each month. Colors of each point represent the month of the year. For ease of comparison, the grey outline on each plot represents the emissions intensities from the other plot.

The hourly MEF (Figure 2b) has less month-to-month variance but is, on average, 50% higher than the corresponding AEF. This is because the marginal peaker plant is most likely coal or diesel – both the most expensive and the least clean source of power in a grid mix (solar PV or wind). There is substantial variation in both intra- and inter-regional variation, suggesting opportunities for temporal load shifting (month and hour) and siting of new loads as mechanisms for reducing the carbon intensity of commercial and industrial activity.

*Wholesale and retail electricity prices*

We provide complementary tariff data by reconciling data from 1,492 industrial and commercial tariff structures. While our past paper reported on 100 industrial tariffs (Chapin et al., 2024), it provided the machine-readable data format that we extend here. In addition to expanded geographic coverage, we provide a data reconciliation method for comparing tariff data structures with other time series data, such as AEF, MEF, and DAM. The data used for analysis in this paper is from 2023 (Chapin et al., 2025a). New data is released publicly on GitHub each month (https://github.com/we3lab/industrial-electricity-tariffs). The Python code for re-collecting the data is published along with the data, so the filter can be customized to modify the query.

Throughout the manuscript, we analyze all 1,492 tariffs regardless of the underlying utility. We considered selecting a default tariff from each utility to avoid biasing the samples towards utilities with more tariff options. However, many utilities provide more and less “aggressive” TOU options, so we felt that selecting a single tariff from the myriad options would bias the samples in a different manner. Additionally, to properly address this bias we would need to know the number of consumers and total consumption under each tariff option, which is not publicly available. As a result, we felt that it was best to analyze all the validated tariffs in order to minimize manipulation of the dataset.

We can categorize tariffs as Seasonal-TOU, Seasonal-NonTOU, Nonseasonal-TOU, or Flat (Chapin et al., 2024). Applying these categorizations to our dataset, 29.2% of tariffs are flat, 50.4% are Seasonal-TOU, 9.2% Nonseasonal-TOU, and 11.2% Seasonal-NonTOU. Visualizations of tariff categorization can be found in Figure S1 (*S.1. Tariff Categorization*).

Analysis of this dataset for 1 MW loads suggests that the average price of electricity from tariffs is higher and more variable in the summer than in the winter. For example, the average energy charge and demand charge in the summer is 11.9 ¢/kWh (Figure 3a) and $9.65/kW (Figure 3c), respectively, as opposed to 11.4 ¢/kWh (Figure 3b) and $8.36/kW in winter (Figure 3d). This is largely due to higher demand for electricity in the summer. Flexibility incentives for tariffs are limited regardless of season, with an average energy charge and demand charge spread of 1.43 and 1.31, respectively, and 95% of tariffs having an energy charge spread of < 6 and a demand charge spread < 15.

Looking more closely at the tariff structure, 50% of the collected tariffs have a single charge tier for both demand and energy (Figure 3e). The second most common tariff structure was a single energy charge tier with no demand charges, representing 15% of the dataset. When multiple charge tiers exist, it is more common to have demand charge tiers than energy charge tiers.

Instead of tariffs, some industrial and commercial power consumers have access to wholesale electricity markets, such as the DAM. Comparing the variation in the peak price premium (i.e., ratio of maximum to minimum price) of electricity (Figure 3f), we see tariffs and DAM prices have distinct hourly variability patterns that may incentivize different behavior by commercial and industrial energy users on the same grid. Of particular note, higher variation in DAM prices do not necessarily correlate with higher variation in tariffs, and negative DAM pricing can incentivize up-flex, while tariffs are currently strictly positive across the U.S.

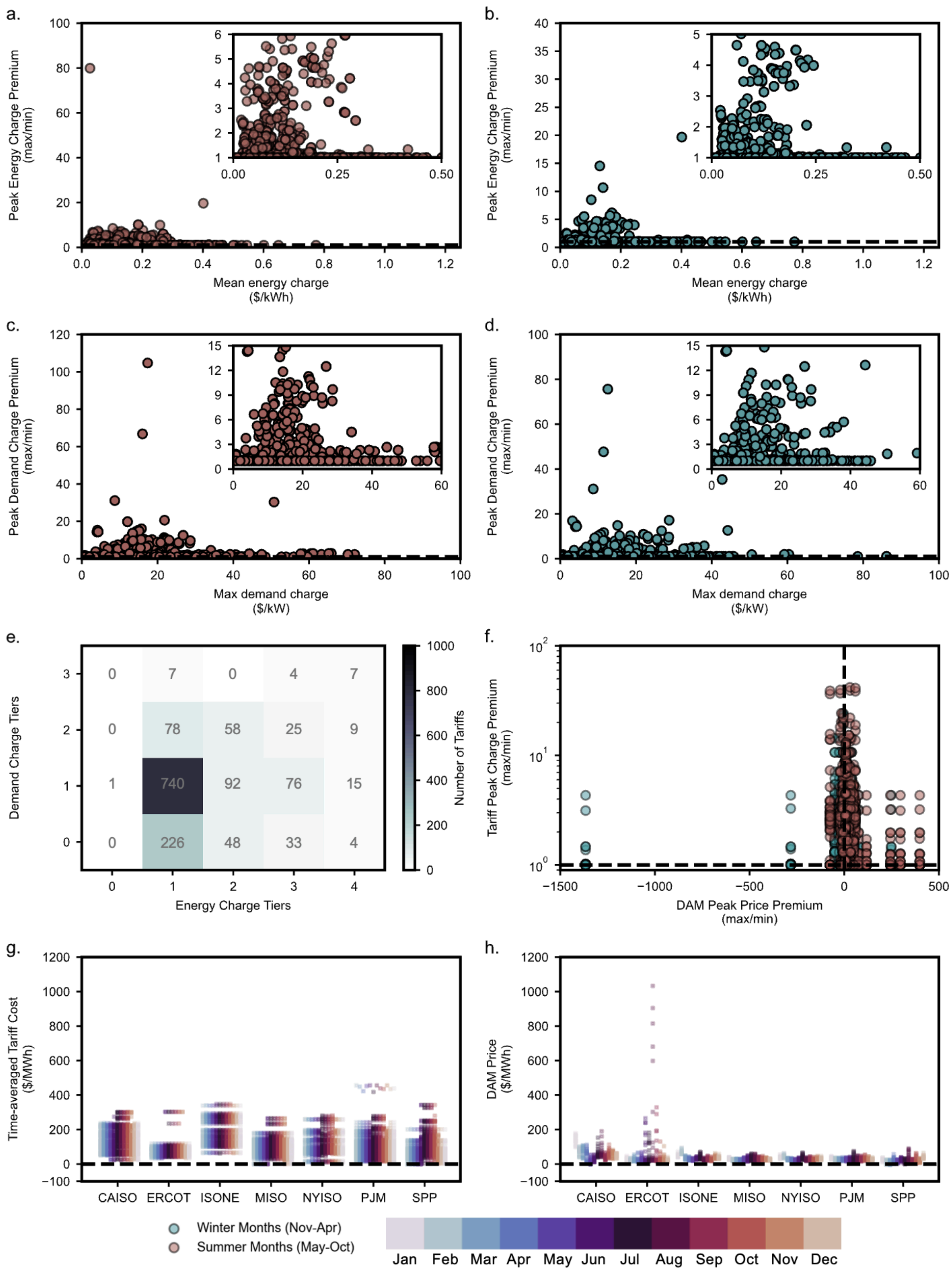

**Figure 3: Charge variation and structure in U.S. retail electricity tariffs in 2023.** a, b) Energy charges (also known as consumption charges) based on kWh of electricity during the Summer (a) and Winter (b). The horizontal axis represents the time-averaged energy charge and the vertical axis represents premium associated with consuming energy during the peak charge period relative to the minimum charge period (i.e., the “Peak Energy Charge Premium”). The inset axes represent the range that covers 95% of the data. c, d) Demand charges (also known as peak power charges) based on the maximum kW of electricity consumed during a specified window during the Summer (c) and Winter (d). The inset axes represent the range that covers 95% of the data. e) Charge tiers are categories of charges triggered based on the amount of use or actual system capacity, with the number inside the heat map representing how many of the 1,492 tariffs fell into that category. 76 tariffs with more than four energy or three demand charge tiers were omitted. f) Peak Price Premium for time-averaged tariff (i.e., both demand and energy charges) versus DAM Peak Price Premium for summer (orange) and winter (teal). g) Time-averaged Tariff Price considers the total average cost of electricity in each ISO/RTO region. The scatter plot shows monthly averages for each region, where the colors of each point represent the month. There are no negative tariffs, but the axis is negative to align with the wholesale price data. h) Day-Ahead Market (DAM) price considers the average price of a MWh of electricity at a given location and time. Values displayed are hourly averages for each month in each ISO/RTO region. Data is obtained directly from Grid Status (GridStatus.io, 2025), MISO (Midcontinent Independent System Operator, 2023), and SPP (Southwest Power Pool, 2023).

Time averaging tariff prices for a 1 MW load reveals strong regional and temporal differences in tariffs (Figure 3g) that are also echoed in the DAM (Figure 3h). There is significant intra- and inter-regional variability within a given month, with up to a 60x variation in the DAM price within an ISO and over 100x difference between some tariffs within the same ISO. Note that the high variation in tariff costs could be due to tariff selection within a utility, which offers different prices based on service voltage, baseload consumption, consumer classification (e.g., industrial, commercial, residential, etc.). Additionally, this analysis is focused on bundled electricity tariffs and most ERCOT tariffs are unbundled (separated into delivery and generation). The Supplementary Information (SI) provides additional analysis of delivery-only charges (*S.1 Tariff Categorization*). Sensitivity analysis of other load shapes reveals that the selection of a flat load was not a significant factor in the results (*S.3.4. Alternative Load Shapes*).

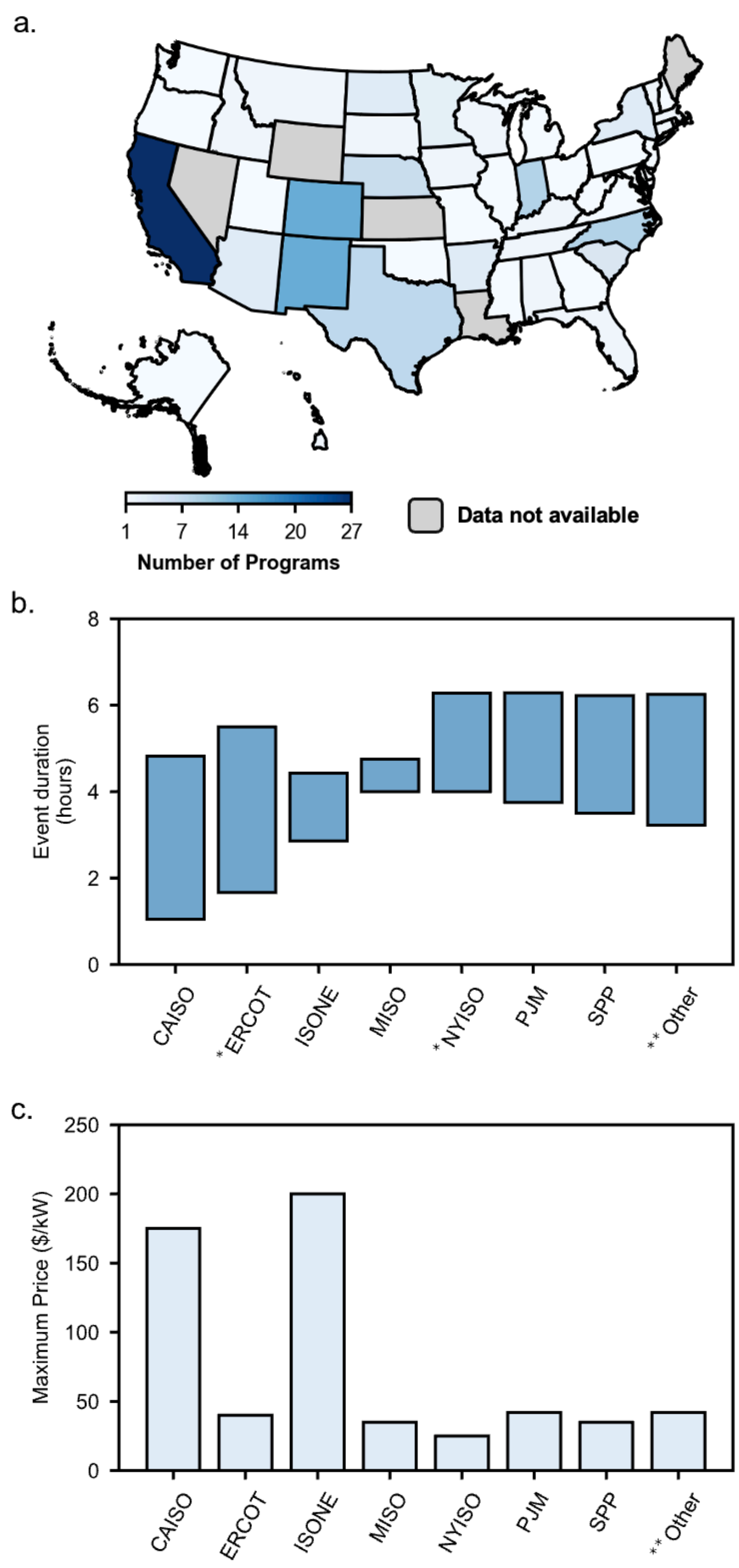

**Figure 4: Incentive-based demand response (IBDR) in 2023.** a) the number of programs per state covered by this dataset are displayed geographically. b) the expected duration of events within programs of each region are shown. The bottom and top of the bar represent the range in possible event durations. c) The maximum payment for programs within this dataset (sorted by ISO/RTO) is shown by the grey bar chart. The payments are estimated based on those programs which have a pre-defined per kW rate that is publicly available. It should be noted that there is a significant intrastate variance in payment structures and amounts, and even more so

when considering demand response aggregators. *When programs have undefined duration limits, we assume a minimum of 1 hour and a maximum of the longest observed event duration (6.3 hours). **Programs classified in the Other region refer to those that do not belong to an ISO.

Finally, we collected, structured, and analyzed national data on DR incentive programs. This dataset has been publicly released on the Stanford Digital Repository as the Incentive Demand Response Program Parameter (IDRoPP) dataset (David et al., 2024). In addition to the subset of program parameters in Table 1, the full description of the IDRoPP data format is available in the SI (*S.2. Incentive-based Demand Response (IBDR) Program Parameters*). Given the prevalence of DR programs in CAISO, this dataset is concentrated in California but has coverage across the U.S. (Figure 4a). Programs in the dataset have event durations centered about the 5-hour range, with ERCOT and CAISO programs on the lower end (Figure 4b). If the average price of electricity in the U.S. is $0.08/kWh, a 1-hour DR event can return 500-2500 hours of electricity consumption (Figure 4c).

<u>Reconciliation of electricity price and emissions datasets</u>

Data reconciliation enables the first national correlations of emission factors and electricity costs. We evaluate trends between AEFs and tariffs, as well as between MEFs and DAM prices. AEF and tariffs are both “smoother” signals for power consumers, thus facilitating planning and design-based analyses. For AEFs, this smoothness stems from month-to-month uniformity in the generation mix being common throughout a month. Tariffs are smooth by design, as they are a set of piecewise linear functions representing the cost of electricity preset by the utility. MEF and DAM are both more volatile and responsive to operational changes than AEF and tariffs. In summary, AEF and tariff are more useful in planning decisions, while MEF and DAM are more useful in operational decisions.

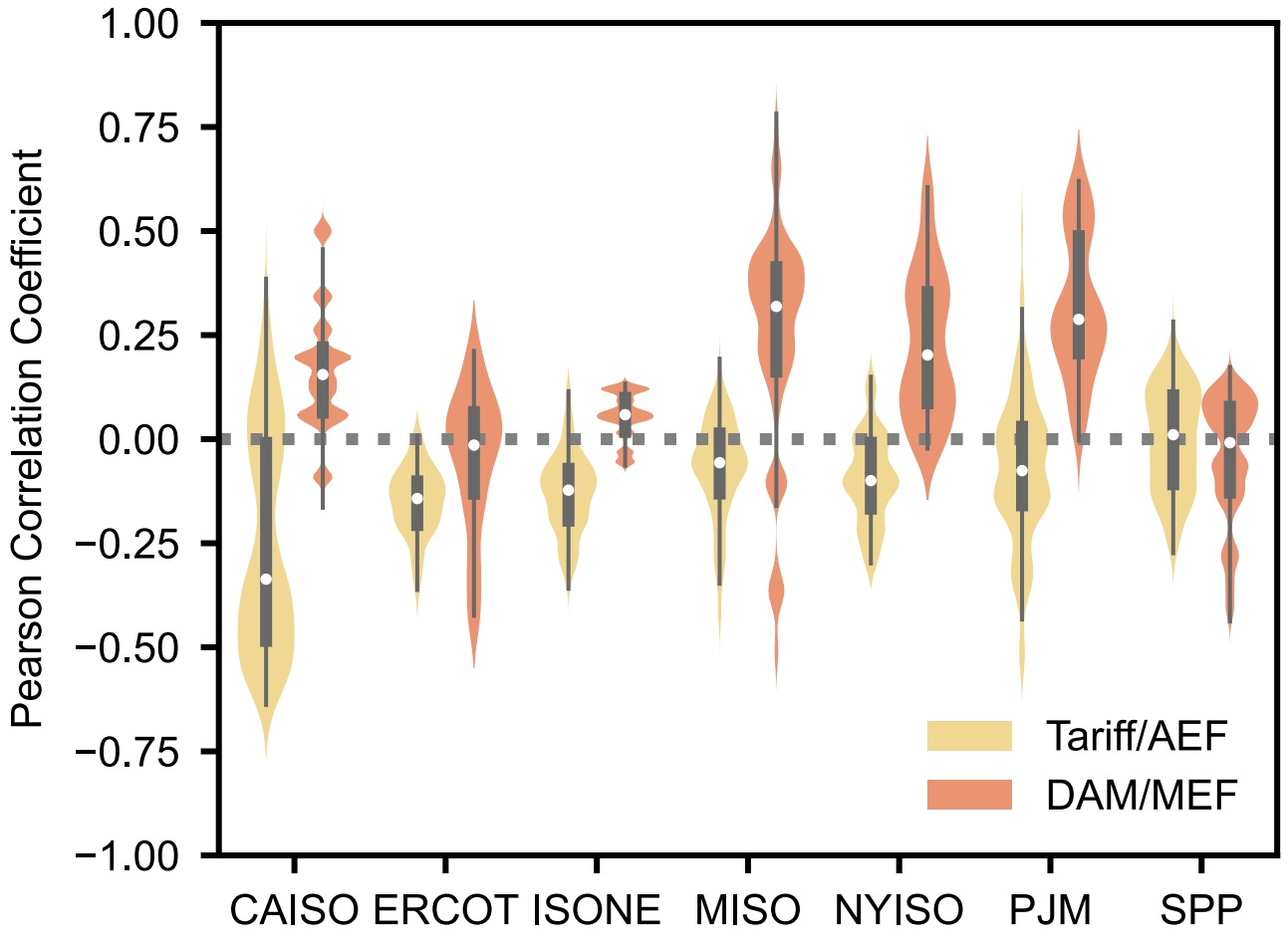

**Figure 5: Correlation of electricity prices and emission intensity in 2023**. Pearson correlation coefficients between historical average emission factors (AEF) and time-averaged

tariffs include both energy and demand charges for each ISO and season pair. Each sample is the Pearson correlation coefficient of a different monthly tariff and the average AEF for the region. Flat tariffs are excluded as they have no correlation. Pearson correlation coefficients of the average marginal emissions factors (MEFs) for a region and day-ahead market (DAM) prices are nodal. Violin plot widths are equal regardless of sample size to highlight the underlying distribution.

On average AEFs and tariffs are negatively correlated, while MEFs and DAM prices are positively correlated (Figure 5). Month-to-month variation is further elucidated in heat maps available in the SI (Figure S3). Unlike DAM prices, retail electricity rates bundle generation, transmission, and distribution costs and account for congestion and resource adequacy requirements. While tariffs more accurately reflect these total system costs, they are also incentivizing electricity use during emissions intensive periods.

We also observe that the correlation distributions for some ISOs are multimodal. There are subclusters of variation across multiple dimensions: temporally (see *S.3.3 Seasonal Variation*), spatially by node (DAM) or utility (tariffs), and consumer classification for tariffs. This multimodal behavior suggests that accurate analysis often requires high spatiotemporal resolution. One exception is the San Francisco Bay Area, where correlation coefficients are seasonally and spatially consistent (Figure S4).

Load shifting incentives for power consumers

Load consumers use price signals, and to a lesser extent, emissions signals, for planning and scheduling commercial and industrial operations. Our initial hypothesis posited that tariffs with high peak-to-off-peak ratios would align retail cost and emissions signals (tariff and AEF) within grid regions. Time-of-use pricing policies reduce peak demand, and we intuited that these policies would reduce the utilization factors of high marginal cost and emissions intensive peaker plants. Contrary to expectations, the analysis revealed no definitive relationship between a tariff's peak-to-off-peak ratio and its correlation with AEFs (Figure 6a). This outcome indicates that elements other than generation, such as distribution and transmission costs, may contribute more substantially to tariff costs in regions where cost and emissions signals are not aligned.

We also hypothesized that higher IBDR payouts (i.e., demand response prices) would occur when costs and emissions are aligned. Our reasoning here was that IBDR programs and tariffs (a form of PBDR) are both policy decisions intended to improve grid operations, so rulemaking bodies may be interested in implementing both IBDR and PBDR. However, IBDR has no apparent relationship to retail costs and emissions signals (Figure 6b). This confirms previous research that suggests DR events are usually called during grid supply deficits and not to avoid reliance on unclean electricity (O'Connell et al., 2014), while also presenting the opportunity for design of an emissions-based demand response program.

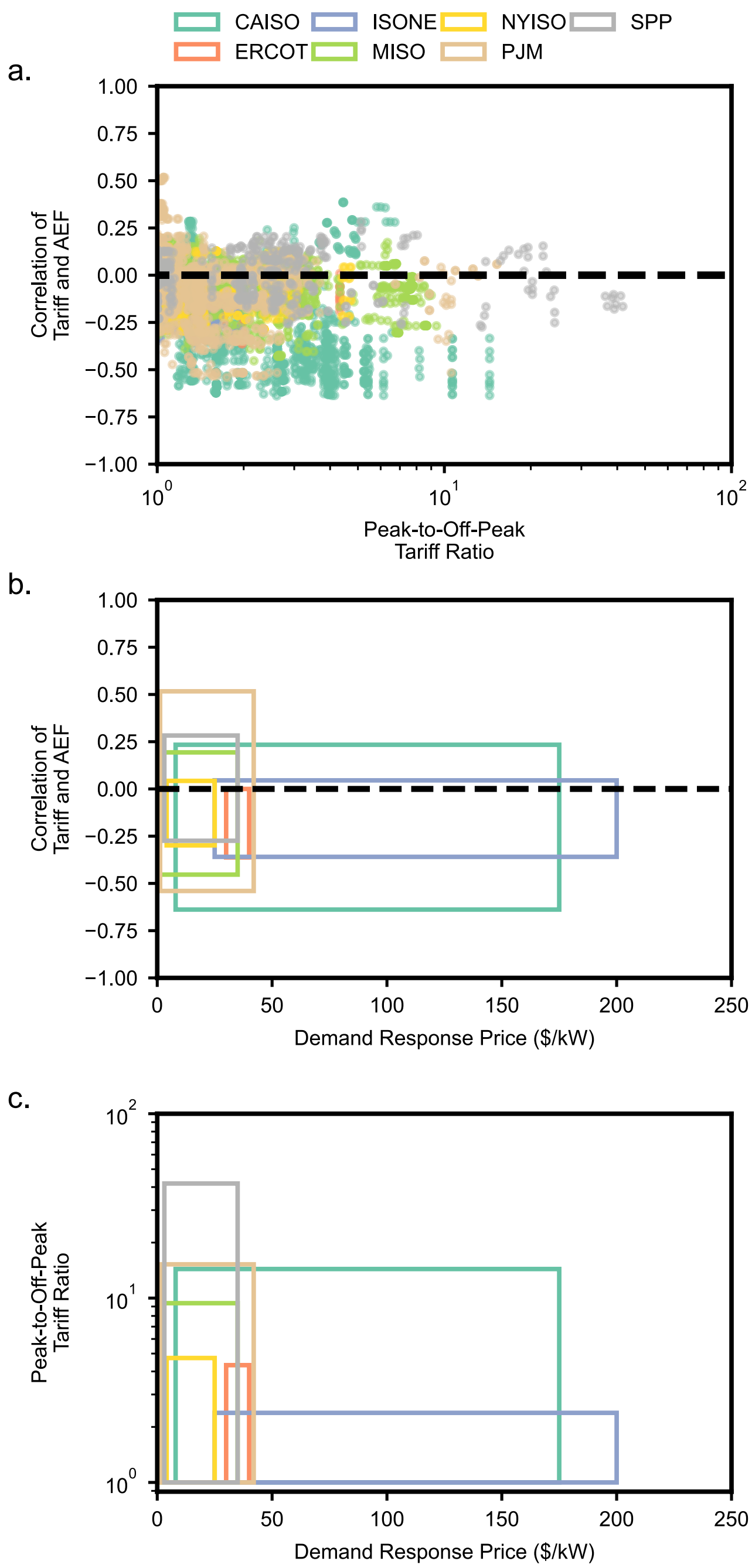

**Figure 6. Program structures and their relation to correlation in 2023.** a) The peak-to-off peak price of electricity tariffs are compared against the timeseries correlation between the associated electricity prices and emissions. b) A spread of demand response prices is compared to the correlation between prices and emissions. Due to the limited number of DR program examples, the spread is shown as a bounding box, rather than individually scattered points. c) A spread of demand response prices is compared to the peak-to-off peak price of electricity tariffs in the same region.

Finally, we hypothesized that regions with high IBDR payouts are likely to have low peak-to-off-peak tariff ratio (i.e., PBDR incentives), and vice versa, because policymakers are likely to incentivize flexibility either through IBDR or PBDR, but not both. This hypothesis holds true for regions other than CAISO, where utilities either have high peak-to-off-peak spreads (e.g., SPP) or high IBDR prices (e.g., ISONE). In CAISO, the tension between PBDR and IBDR program incentives encourages load consumers to increase power consumption during non-DR peak hours to drive up their baseline power consumption and increase their payout. This is a suboptimal result for both the grid, increasing demand during peak hours, and the industrial load consumer that must pay the difference on increased peak electricity charges. The tension between IBDR and PBDR is well documented in prior research (Dehnavi and Abdi, 2016), but has not been so clearly analyzed at a national policy level.

## Discussion

This paper provides the first nationwide analysis of alignment between cost and emission incentives for commercial and industrial power consumers. First, we supplement previously published emissions and wholesale electricity cost datasets by collecting nationwide incentive-based demand response (IBDR) program and retail electricity (i.e., tariff) data. Second, we reconcile incompatibilities in data structure and temporal and geospatial data fragmentation to enable combined analysis of the datasets. Third, we analyze the reconciled data to investigate the alignment of emissions and cost incentives across the United States.

The nationwide scope of this analysis facilitates comparison of grid regions with diverse structural and policy features. While this feature heterogeneity inhibits rigorous causal analyses for data collected in a single year, it facilitates observational analysis of trends that can be more rigorously tested in future work. One such trend is the degree to which commercial and industrial load consumers are subject to meaningful price or emissions incentives for load shifting. For 2023, we observe the largest price-based variation in ERCOT (DAM prices) and NYISO, CAISO, and ISONE (tariffs). The value of commercial and industrial load shifting was either very low in other regions or is poorly captured in current tariff structures, which is consistent with other work (Rao et al., Under review). This inter- and intraregional variation in electricity prices stems from spatiotemporal differences in generation mix and market-based dispatch, climate-driven demand profiles, costs associated with transmission and distribution, and policy decisions such as the appropriate complexity of tariff structures. In contrast, the hourly variability of grid emissions is greatest in PJM (MEFs) and SPP (AEFs), with ERCOT and CAISO also featuring significant variation in AEFs. Emissions factor variance is also caused by diversity in the generation mix and the order of generator dispatch, which is a function of power demand, fuel cost, and local congestion. High variance signals a market opportunity for flexible power consumers to reduce emissions by scheduling consumption.

The structural and policy origins of this nationwide variation in price-based and emissions incentives are also reflected in the degree of alignment between incentives. As highlighted above, MEF and DAM are positively correlated for most seasons and regions. This finding supports past research proposing the use of wholesale prices as a proxy for MEF (He et al., 2024). In addition, the new data reconciliation methods presented in this work enable novel analysis of alignment between tariffs and Scope 2 emissions with higher spatiotemporal resolution than past work (Kaufmann and Vaid, 2016). On average, we observe negative correlations between AEF and tariffs. In other words, an economically rational, flexible retail electricity consumer is more likely to increase, rather than decrease, grid emissions.

The origins of this misalignment in cost and carbon incentives for retail customers are ISO region specific and will likely require a diverse set of market and policy tools to correct. Locations with substantial variation in emissions factors and a negative correlation between emissions and cost signals could adopt tools for incorporating a price on carbon (e.g., carbon tax, carbon market, or higher renewable energy credit requirements), implement emissions-focused demand response markets, or intentionally design tariffs to align with emissions factors to help correct this misalignment of incentives. Addressing this misalignment would incentivize commercial and industrial loads to reduce both power costs and emissions via flexible operation.

In addition to documenting variability in the magnitude and alignment of incentives for load flexibility, the proposed methods for nationwide analysis could inform the siting of large new commercial and industrial loads or investments in energy flexibility upgrades. While there are general trends at the ISO level, as shown by Figure 4, there is significant variability in alignment at the nodal level. Proximal nodes may exhibit similar behavior (Figure S4), but ISOs contain wide-spanning networks. Accurate load siting will require nodal information, as well as a detailed understanding of the flexibility characteristics of the commercial or industrial load. The analysis conducted in this manuscript assumes a perfectly rational, totally flexible electricity consumer, but real-world consumers are constrained. This raises the question of how these constraints, or flexibility characteristics, influence the potential to realize emissions reductions.

These flexibility characteristics, including the ability to vary power consumption continuously or operate intermittently, determine how commercial and industrial loads exploit regional cost and emissions arbitrage opportunities (Rao et al., 2024). Steady state loads minimize costs and emissions by siting in low average cost states with aligned emissions signals. Conversely, highly flexible loads are most beneficially sited in regions with the highest peak-to-off-peak ratio that is emissions-aligned to take advantage of flexible operation. Finally, minimally operational flexibility loads that can completely or partially shut down are optimally located in regions with the highest IBDR payouts. While the cost and emissions intensity of production are rarely the sole factor in siting new loads or enhancing the flexibility of existing ones, the methods presented here may be integrated as part of the design process, alongside other considerations such as capital and operational expenses.

Finally, the methods presented here are relevant to designing tariffs and IBDR programs that economically incentivize emissions reductions. The new datasets could be merged with additional tariff datasets (Smart Electric Power Alliance, 2025) or integrated with other tariff-design approaches (Shi and Xu, 2024) to more broadly represent emerging markets. Using our method to quantify alignment before a tariff is published could ensure that new electricity tariffs are emissions aligned. Future work could also build upon this data reconciliation method to facilitate probabilistic analysis required for fully capturing the value of IBDR participation. As previously discussed, the expected payout from IBDR events is stochastic and highly dependent on exogenous factors (from a consumer perspective), such as when IBDR events are called and how the baselining procedure is conducted. IBDR programs could be accurately modeled by quantifying the uncertainties relative to underlying TOU tariffs, as well as the alignment of IBDR programs and emissions signals.

**Acknowledgements**
This work was supported by the following grants and programs: National Alliance for Water Innovation (NAWI) (grant number UBJQH); Department of Energy, the Office of Energy Efficiency and Renewable Energy, Advanced Manufacturing Office (grant number DE-EE0009499); California Energy Commission (CEC) (grant number GFO-23-316); Equitable, Affordable & Resilient Nationwide Energy System Transition (EARNEST) Consortium; Stanford University Bits & Watts Initiative; Stanford Woods Institute Realizing Environmental Innovation Program (REIP); Stanford Woods Institute Mentoring Undergraduate in Interdisciplinary Research (MUIR) Program; and Stanford University Sustainability Undergraduate Research in Geoscience and Engineering (SURGE) Program. The authors would like to thank Kamran Tehranchi and Ines Azevedo from Stanford University for helpful conversations during the drafting process.

**Author Contributions**
**F.T.C.** contributed to conceptualization, data curation, formal analysis, investigation, methodology, project administration, software, validation, visualization, writing (original draft), and writing (review & editing).
**A.K.R.** contributed to conceptualization, data curation, formal analysis, investigation, methodology, project administration, software, validation, visualization, writing (original draft), and writing (review & editing).
**A.S.** contributed to conceptualization, data curation, formal analysis, investigation, methodology, software, writing (original draft), and writing (review & editing).
**C.I.T.** contributed to conceptualization, data curation, writing (original draft), and writing (review & editing).
**E.D.** contributed to data curation, visualization, and writing (original draft).
**C.S.C.** contributed to data curation, software, and writing (original draft).
**E.M.** contributed to conceptualization and writing (review & editing).
**M.S.M.** contributed to conceptualization, funding acquisition, project administration, supervision, validation, and writing (review & editing).

## References

Ahlers, D., 2013. Assessment of the accuracy of GeoNames gazetteer data, in: Proceedings of the 7th Workshop on Geographic Information Retrieval. Presented at the SIGSPATIAL'13: 21st SIGSPATIAL International Conference on Advances in Geographic Information Systems, ACM, Orlando Florida, pp. 74–81. https://doi.org/10.1145/2533888.2533938

Bellinguer, K., Girard, R., Bocquet, A., Chevalier, A., 2023. ELMAS: a one-year dataset of hourly electrical load profiles from 424 French industrial and tertiary sectors. Sci Data 10, 686. https://doi.org/10.1038/s41597-023-02542-z

Blumsack, S., 2009. Electric rate design and greenhouse-gas emissions reduction, in: 2009 IEEE Power & Energy Society General Meeting. Presented at the IEEE Power & Energy Society General Meeting, IEEE, Calgary, AB, Canada, pp. 1–4. https://doi.org/10.1109/PES.2009.5275332

Borenstein, S., Kellogg, R., 2023. Carbon Pricing, Clean Electricity Standards, and Clean Electricity Subsidies on the Path to Zero Emissions. Environmental and Energy Policy and the Economy 4, 125–176. https://doi.org/10.1086/722675

Callaway, D.S., Fowlie, M., McCormick, G., 2018. Location, Location, Location: The Variable Value of Renewable Energy and Demand-Side Efficiency Resources. Journal of the Association of Environmental and Resource Economists 5, 39–75. https://doi.org/10.1086/694179

Chapin, F.T., Bolorinos, J., Mauter, M.S., 2024. Electricity and natural gas tariffs at United States wastewater treatment plants. Sci Data 11, 113. https://doi.org/10.1038/s41597-023-02886-6

Chapin, F.T., Rao, A.K., Sakthivelu, A., Chen, C.S., Mauter, M.S., 2025a. Industrial and Commercial Electricity Tariffs in the United States. https://doi.org/10.5281/ZENODO.16739989

Chapin, F.T., Rao, A.K., Sakthivelu, A., Wettermark, D., Jaminet, A., Dudchenko, A.V., Mauter, M., 2025b. Electric Emissions & Cost Optimizer (EECO). https://doi.org/10.5281/ZENODO.17102024

David, E., Sakthivelu, A., Rao, A.K., Mauter, M.S., 2024. US incentive based demand response program parameters. https://doi.org/10.25740/ck480bd0124

de Chalendar, J.A., Benson, S.M., 2021. A physics-informed data reconciliation framework for real-time electricity and emissions tracking. Appl. Energy 304, 117761. https://doi.org/10.1016/j.apenergy.2021.117761

de Chalendar, J.A., Benson, S.M., 2019. Why 100% Renewable Energy Is Not Enough. Joule 3, 1389–1393. https://doi.org/10.1016/j.joule.2019.05.002

de Chalendar, J.A., Taggart, J., Benson, S.M., 2019. Tracking emissions in the US electricity system. Proc. Natl. Acad. Sci. U.S.A. 116, 25497–25502. https://doi.org/10.1073/pnas.1912950116

Dehnavi, E., Abdi, H., 2016. Optimal pricing in time of use demand response by integrating with dynamic economic dispatch problem. Energy 109, 1086–1094. https://doi.org/10.1016/j.energy.2016.05.024

Evro, S., Alamooti, M., Tomomewo, O.S., 2026. Quantifying the global energy transition: A policy-ready framework linking renewable deployment and emissions outcomes.

Renewable and Sustainable Energy Reviews 225, 116189. https://doi.org/10.1016/j.rser.2025.116189
Gorka, J., Rhodes, N., Roald, L., 2025. ElectricityEmissions.jl: A Framework for the Comparison of Carbon Intensity Signals, in: Proceedings of the 16th ACM International Conference on Future and Sustainable Energy Systems. Presented at the E-Energy ’25: The 16th ACM International Conference on Future and Sustainable Energy Systems, ACM, Rotterdam Netherlands, pp. 19–30. https://doi.org/10.1145/3679240.3734597
Goteti, N.S., Hittinger, E., Sergi, B., Lima Azevedo, I., 2021. How does new energy storage affect the operation and revenue of existing generation? Applied Energy 285, 116383. https://doi.org/10.1016/j.apenergy.2020.116383
GridStatus.io, 2025. Grid Status [WWW Document]. URL https://www.gridstatus.io/ (accessed 5.12.25).
Harris, C.R., Millman, K.J., Van Der Walt, S.J., Gommers, R., Virtanen, P., Cournapeau, D., Wieser, E., Taylor, J., Berg, S., Smith, N.J., Kern, R., Picus, M., Hoyer, S., Van Kerkwijk, M.H., Brett, M., Haldane, A., Del Río, J.F., Wiebe, M., Peterson, P., Gérard-Marchant, P., Sheppard, K., Reddy, T., Weckesser, W., Abbasi, H., Gohlke, C., Oliphant, T.E., 2020. Array programming with NumPy. Nature 585, 357–362. https://doi.org/10.1038/s41586-020-2649-2
He, X., Tsang, D.H.K., Chen, Y., 2024. Is Locational Marginal Price All You Need for Locational Marginal Emission? https://doi.org/10.48550/ARXIV.2411.12104
Huggins, J., 2024. U.S. Electric Utility Companies and Rates: Look-up by Zipcode (2023).
Jiang, W., Huber, O., Ferris, M.C., Roald, L., 2025. Can Carbon-Aware Electric Load Shifting Reduce Emissions? An Equilibrium-Based Analysis. https://doi.org/10.48550/arXiv.2504.07248
Kaufmann, R.K., Vaid, D., 2016. Lower electricity prices and greenhouse gas emissions due to rooftop solar: empirical results for Massachusetts. Energy Policy 93, 345–352. https://doi.org/10.1016/j.enpol.2016.03.006
Kök, A.G., Shang, K., Yücel, Ş., 2018. Impact of Electricity Pricing Policies on Renewable Energy Investments and Carbon Emissions. Management Science 64, 131–148. https://doi.org/10.1287/mnsc.2016.2576
Lin, L., Zavala, V.M., Chien, A.A., 2021. Evaluating Coupling Models for Cloud Datacenters and Power Grids, in: Proceedings of the Twelfth ACM International Conference on Future Energy Systems. Presented at the e-Energy ’21: The Twelfth ACM International Conference on Future Energy Systems, ACM, Virtual Event Italy, pp. 171–184. https://doi.org/10.1145/3447555.3464868
Mayes, S., Sanders, K., 2022. Quantifying the electricity, $CO_2$ emissions, and economic tradeoffs of precooling strategies for a single-family home in Southern California*. Environ. Res.: Infrastruct. Sustain. 2, 025001. https://doi.org/10.1088/2634-4505/ac5d60
Mayes, S., Zhang, T., Sanders, K.T., 2024. Residential precooling on a high-solar grid: impacts on $CO_2$ emissions, peak period demand, and electricity costs across California. Environ. Res.: Energy 1, 015001. https://doi.org/10.1088/2753-3751/acfa91

McLaren, J., Gagnon, P., Zimny-Schmitt, D., DeMinco, M., Wilson, E., 2017. Maximum demand charge rates for commercial and industrial electricity tariffs in the United States. https://doi.org/10.7799/1392982
Midcontinent Independent System Operator, 2023. Day-Ahead Pricing (xls) [WWW Document]. Market Reports. URL https://www.misoenergy.org/markets-and-operations/real-time--market-data/market-reports/ (accessed 8.25.25).
Neutel, J., Berson, A., Saltzer, S., Brandt, A., Weyant, J., Orr, F.M., Benson, S.M., 2026. What will it take to get to net-zero emissions in California? Energy Policy 208, 114848. https://doi.org/10.1016/j.enpol.2025.114848
O'Connell, N., Pinson, P., Madsen, H., O׳Malley, M., 2014. Benefits and challenges of electrical demand response: A critical review. Renewable and Sustainable Energy Reviews 39, 686–699. https://doi.org/10.1016/j.rser.2014.07.098
OpenEI, 2025. United States Utility Rate Database (USURDB).
pgeocode v0.5.0, 2024. https://pypi.org/project/pgeocode/
Puschnigg, S., Knöttner, S., Lindorfer, J., Kienberger, T., 2023. Development of the virtual battery concept in the paper industry: Applying a dynamic life cycle assessment approach. Sustainable Production and Consumption 40, 438–457. https://doi.org/10.1016/j.spc.2023.07.013
Rao, A., K., Chapin, F.T., Musabandesu, E., Sakthivelu, A., Tucker, C.I., Wettermark, D., Mauter, M.S., Under review. How much can we save? Upper bound cost and emissions benefits from commercial and industrial load flexibility.
Rao, A.K., Bolorinos, J., Musabandesu, E., Chapin, F.T., Mauter, M.S., 2024. Valuing energy flexibility from water systems. Nat Water 2, 1028–1037. https://doi.org/10.1038/s44221-024-00316-4
Ricks, W., Xu, Q., Jenkins, J.D., 2023. Minimizing emissions from grid-based hydrogen production in the United States. Environ. Res. Lett. 18, 014025. https://doi.org/10.1088/1748-9326/acacb5
Sakthivelu, A., Chapin, F.T., Bolorinos, J., Mauter, M.S., 2025. Demand Response Event Simulator and Risk-Aware Bidding Tool for Industrial Customers. Manuscript submitted for publication.
Shi, Y., Xu, B., 2024. Demand-side price-responsive flexibility and baseline estimation through end-to-end learning. IET Renewable Power Gen 18, 361–371. https://doi.org/10.1049/rpg2.12794
Siler-Evans, K., Azevedo, I.L., Morgan, M.G., 2012. Marginal Emissions Factors for the U.S. Electricity System. Environ. Sci. Technol. 46, 4742–4748. https://doi.org/10.1021/es300145v
Smart Electric Power Alliance, 2025. Database of Emerging Large-Load Tariffs (DELTa).
Sofia, S., Dvorkin, Y., 2024. Carbon Impact of Intra-Regional Transmission Congestion. https://doi.org/10.2139/ssrn.4972564
Sonali Razdan, Jennifer Downing, Louise White, 2025. Pathways to Commercial Liftoff: Virtual Power Plants 2025 Update. US Department of Energy.
Southwest Power Pool, 2023. Day-Ahead Market [WWW Document]. SPP Portal. URL https://portal.spp.org/groups/day-ahead-market (accessed 8.25.25).

Suri, D., Chalendar, J. de, Azevedo, I., 2024. What are the real implications for CO2 as generation from renewables increases? https://doi.org/10.48550/arXiv.2408.05209
Tamayao, M.-A.M., Michalek, J.J., Hendrickson, C., Azevedo, I.M.L., 2015. Regional Variability and Uncertainty of Electric Vehicle Life Cycle $CO_2$ Emissions across the United States. Environ. Sci. Technol. 49, 8844–8855. https://doi.org/10.1021/acs.est.5b00815
U.S. Department of Energy, 2024a. Demand Response and Time-Variable Pricing Programs [WWW Document]. Federal Energy Management Program. URL https://www.energy.gov/femp/demand-response-and-time-variable-pricing-programs (accessed 6.1.24).
U.S. Department of Energy, 2024b. Demand Response and Time-Variable Pricing Programs Search [WWW Document]. Federal Energy Management Program. URL https://www.energy.gov/femp/demand-response-and-time-variable-pricing-programs-search (accessed 6.1.24).
U.S. Department of Homeland Security, 2025. Homeland Infrastructure Foundation-Level Data.
U.S. Energy Information Administration, 2023. Hourly Electric Grid Monitor [WWW Document]. EIA Open Data. URL https://www.eia.gov/electricity/gridmonitor/about (accessed 7.21.24).
U.S. Environmental Protection Agency, 2023. Clean Air Markets Program Data: Part 75 Emissions Data.
WattTime, 2025. Home - WattTime [WWW Document]. URL https://www.watttime.org/ (accessed 5.12.25).
Wei, M., Nelson, J.H., Greenblatt, J.B., Mileva, A., Johnston, J., Ting, M., Yang, C., Jones, C., McMahon, J.E., Kammen, D.M., 2013. Deep carbon reductions in California require electrification and integration across economic sectors. Environ. Res. Lett. 8, 014038. https://doi.org/10.1088/1748-9326/8/1/014038
Wick, M., 2025. GeoNames [WWW Document]. GeoNames. URL https://www.geonames.org/ (accessed 8.25.25).
Zhou, E., Mai, T., 2021. Electrification Futures Study: Operational Analysis of U.S. Power Systems with Increased Electrification and Demand-Side Flexibility (No. NREL/TP--6A20-79094, 1785329, MainId:33320). https://doi.org/10.2172/1785329

**Supplementary Information for "Retail electricity costs and emissions incentives are misaligned for commercial and industrial power consumers"**

*Fletcher T. Chapin[a], Akshay K. Rao[a], Adhithyan Sakthivelu[a], Carson I. Tucker[b], Eres David[c], Casey S. Chen[d], Erin Musabandesu[a], Meagan S. Mauter[a,e,f,g,h,*]*

[a]Department of Civil and Environmental Engineering, Stanford University, 473 Via Ortega, Stanford, California 94305, United States
[b]Department of Mechanical Engineering, Stanford University, 440 Escondido Mall Building 530, Stanford, California 94305, United States
[c]Department of Earth and Environmental Engineering, Columbia University, 500 W 120th St, New York, New York 10027, United States
[d]Symbolic Systems Program, Stanford University, 389 Jane Stanford Way, Stanford, California 94305, United States
[e]Department of Environmental Social Sciences, Stanford University, 473 Via Ortega, Stanford, California 94305, United States
[f]Senior Fellow, Woods Institute for the Environment, Stanford University, 473 Via Ortega, Stanford, California 94305, United States
[g]Senior Fellow, Precourt Institute for Energy, Stanford University, 473 Via Ortega, Stanford, California 94305, United States
[h]Photon Science, SLAC National Accelerator Laboratory, 2575 Sand Hill Road, Menlo Park, California 94025, United States

*Author contact: mauter@stanford.edu

*Number of pages: 15*
*Number of tables: 1*
*Number of figures: 7*

# S.1. Tariff Categorization

## S.1.1. Temporality

Using the classification schema from Chapin et al. (Chapin et al., 2024), we categorize monthly electricity energy and demand charges into Flat, Seasonal-TOU, Nonseasonal-TOU, or Seasonal-NonTOU charges for the 1,492 bundled tariffs analyzed in the main manuscript (Figure S1):

- *Flat*: Charges are constant throughout the year.
- *Seasonal-TOU*: Charges vary monthly (Seasonal) and daily and/or hourly (TOU).
- *Nonseasonal-TOU*: Charges are consistent month-to-month but vary daily and/or hourly.
- *Seasonal-NonTOU*: Charges vary monthly but constant from day to day and hour to hour.

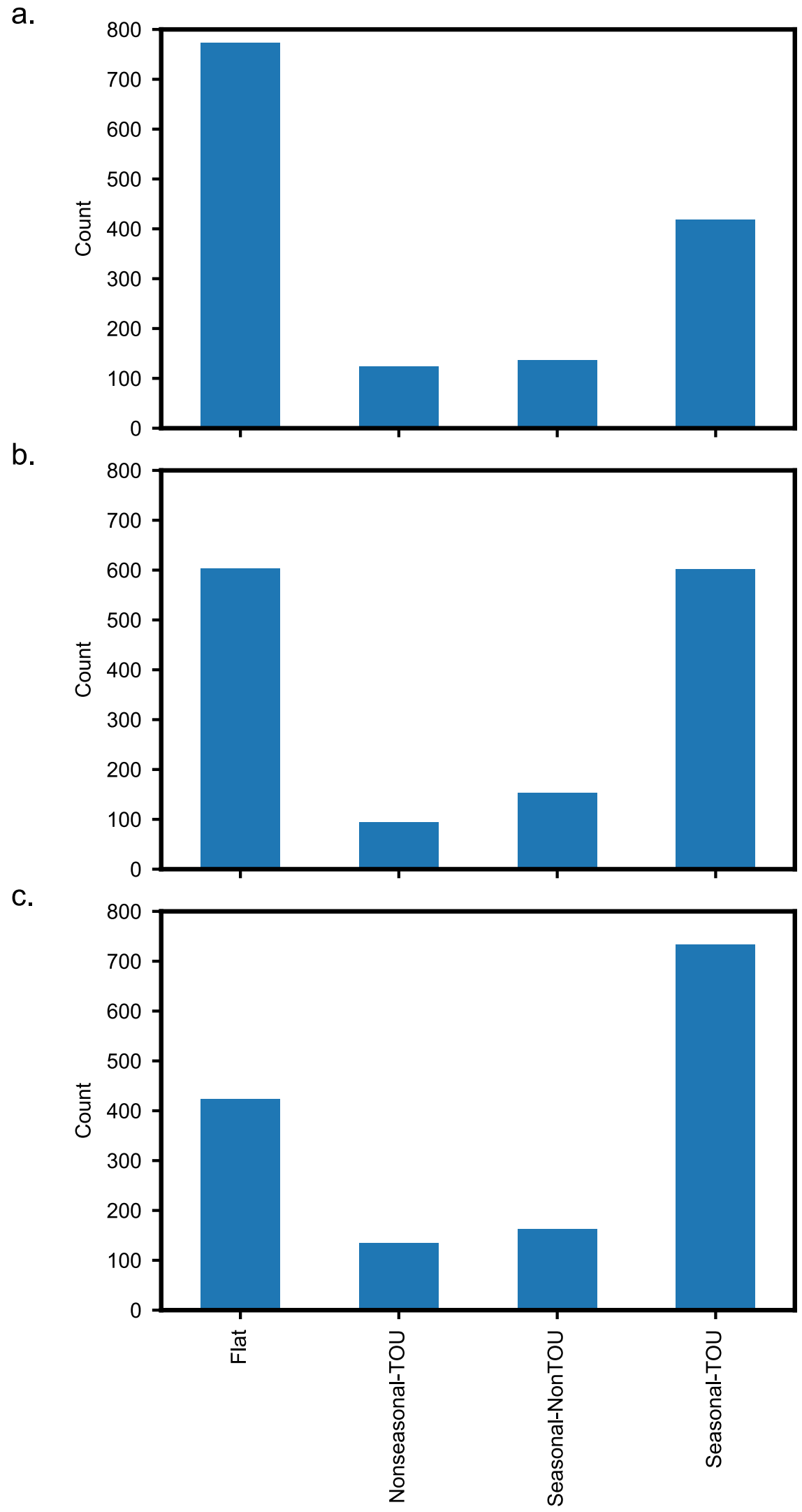

**Figure S1. Categorization of bundled US electricity tariffs in 2023.** Tariffs are categorized as no variation (“Flat”), month-to-month and hourly variation (“SeasonalTOU”), hourly but no month-to-month variation (“NonseasonalTOU”), and monthly and hourly variation (“SeasonalNonTOU”) for a) energy charges, b) demand charges, and c) overall tariff (i.e., both energy and demand charges).

## S.1.2. Bundling

In the United States, generation, transmission, and distribution charges are commonly bundled into a single electricity tariff. However, in some cases, facilities pay separate generation (or supply) and delivery (or transmission and distribution) charges to a single provider or to separate generation and delivery providers. To address this discrepancy, we focus the main manuscript on bundled tariffs and conduct a limited analysis of delivery-only tariffs in the Supplementary Information (Figure S2).

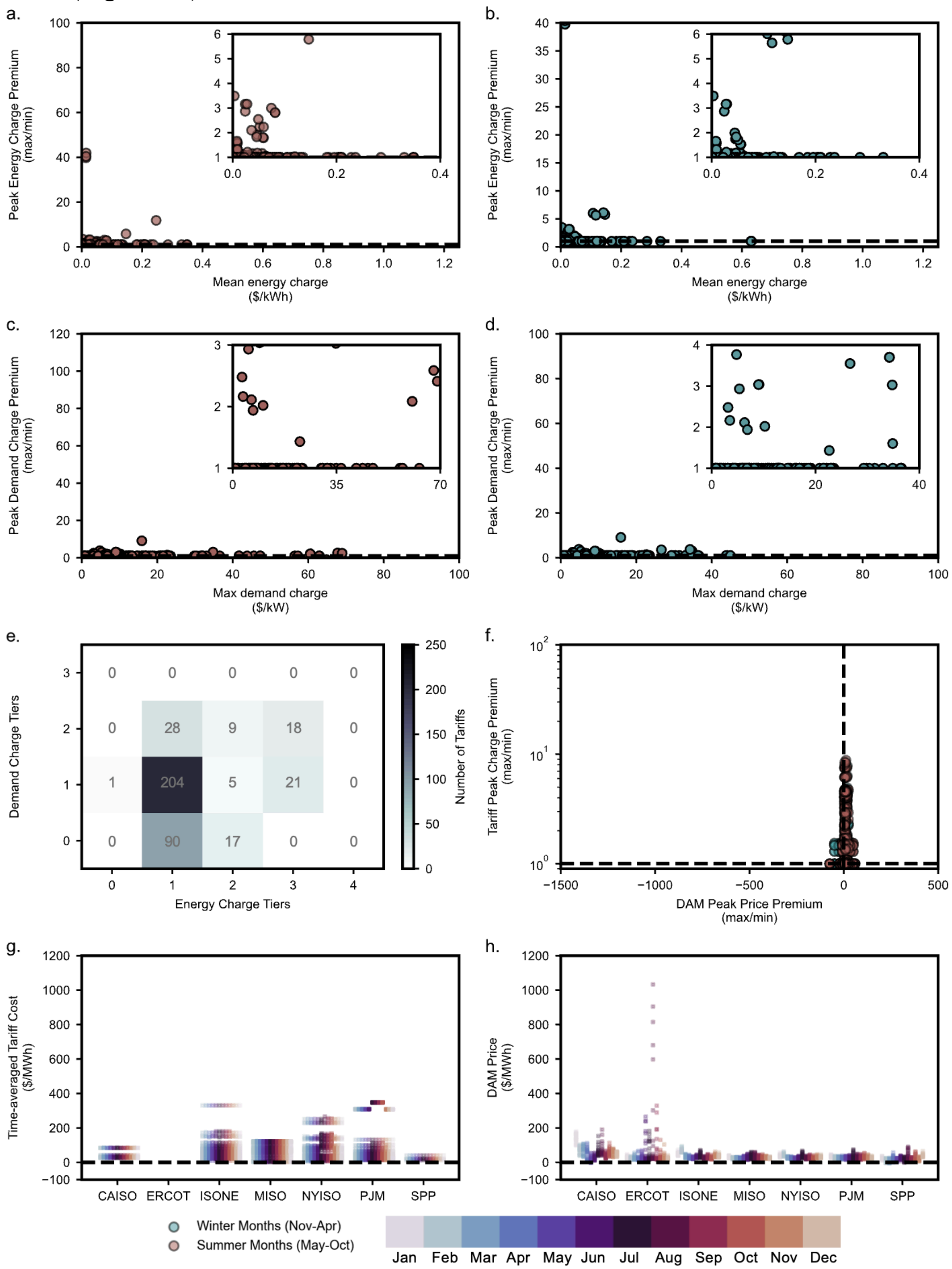

**Figure S2. Charge variation and structure in delivery-only US retail electricity tariffs in 2023.** a, b) Energy charges (also known as consumption charges) based on kWh of electricity during the Summer (a) and Winter (b). The horizontal axis represents the time-averaged energy charge, and the vertical axis represents the premium associated with consuming energy during the peak charge period relative to the minimum charge period (i.e., the "Peak Energy Charge Premium"). The inset axes represent the range that covers 95% of the data. c, d) Demand charges (also known as peak power charges) based on the maximum kW of electricity consumed during a specified window during the Summer (c) and Winter (d). The inset axes represent the range that covers 95% of the data. e) Charge tiers are categories of charges triggered based on the amount of use or actual system capacity, with the number inside the heat map representing how many of the 392 tariffs fell into each category. f) Time-averaged Tariff Price considers the total average cost of electricity in each ISO/RTO region. The scatter plot shows monthly averages for each region, where the colors of each point represent the month. There are no negative tariffs, but the axis is negative to align with the wholesale price data. g) Day-Ahead Market (DAM) price considers the average price of a MWh of electricity at a given location and time. Values displayed are hourly averages for each month in each ISO/RTO region. Data is obtained directly from Grid Status (GridStatus.io, 2025), MISO (Midcontinent Independent System Operator, 2023), and SPP (Southwest Power Pool, 2023).

## S.2. Incentive-based Demand Response (IBDR) Program Parameters

Data in the Incentive Demand Response Program Parameter (IDRoPP) dataset consists of metadata (1 CSV file) and program data (1 CSV file):

- `program_parameters.csv` contains the metadata
- `us_program_parameters.csv` contains the program data

The dataset can be found in the Stanford Digital Repository (David et al., 2024).

### S.2.1. Metadata

Metadata is stored in two CSV files, one with program parameters and one with simulation parameters. Each parameter takes up one row. Table S1 contains a full list of attributes used in the dataset.

**Table S1. Full description of Incentive Demand Response Program Parameter (IDRoPP) dataset attributes.** "Column name" is the plain language shorthand, "Column ID" is the programmatic variable name, and "Column description" is a full-length description of the column data.

| Column name | Column ID | Column description |
|---|---|---|
| Minimum number of event days | min_days | The minimum number of days the program event is called for a customer per month |
| Maximum number of event days | max_days | The maximum number of days the program event is called for a customer per month |
| Minimum duration of event | min_dur | The event should last for more than the minimum duration specified by the program |
| Maximum duration of event | max_dur | The event should not last longer than the maximum duration specified by the program |
| Program start time | start_time | The program can be either 24 hours or last for a specified period of time, and the start time is generally provided if it is not a 24-hour period |
| Program end time | end_time | The program can be either 24 hours or last for a specified period of time, and the end time is generally provided if it is not a 24-hour period |
| Maximum events | max_events | Maximum number of events that can be called in a season |
| Maximum event hours | max_hours | Maximum total hours of all events called in a year |
| Events per day | events_daily | The maximum event that a customer can provide on a single day |
| Maximum consecutive event days | max_consec | The maximum consecutive days the customer can be called in a particular month |
| Notification type | notif_type | The event is generally notified the day before or the day of and is captures by this parameter |
| Notification time | notif_time | If the event is notified the day before or the day of the program generally specifies the time. *Note*: This can also be historic event related |
| Notification time delta | notif_delt | The number of hours between the notification time and the beginning of an event |
| Baseline calculation method | base_method | The method used by each program to calculate the baseline energy usage |
| Presence of historic data | hist_pres | Programs may or may not have historic data present |
| Payment function | pay_function | Form of compensation for shifts in energy load (or being available to shift energy loads) when called upon |
| Region | region | The US DOE separates the states into West Region, The Southeast and Midwest Region, and Northeast Region |
| Day of the week | dow | The days of week on which events can fall |
| Season | season | Some programs run during the summer, winter, or both |
| Eligibility | elig | Eligibility to participate in the program (minimum bids, minimum peak demand, etc.) |
| Company | comp | Which company or companies offer this program |
| Start month | sm | Month during which the DR event season starts |
| End month | em | Month during which the DR event season ends |

| State | state | State in which the program is eligible |
|---|---|---|
| Utility | util | Eligible utilities |
| Trigger | trigger | Event trigger type |
| Eligible load type | load | What type of load is eligible for participation in the program |
| Program or rate | program_rate | Classification as a program or a rate by the DOE |
| Baseline Function | function_base | Function for calculating payment after an event |
| Delivered Ratio | delivered_ratio | The amount reduced divided by nomination of delivery |
| Amount reduced | amount_reduced | Baseline minus consumption |
| Inclusion of weekends | weekends | If the baseline calculation includes weekends |
| Inclusion of holidays | holidays | If the baseline calculation includes holidays |
| Inclusion of previous events | prev_events | If the baseline calculation includes previous events |
| Baseline hours | base_hours | Time of day during which load measurements can be taken for baseline calculation |
| Range value | range_val | Number of load measurements taken at a certain frequency for baseline calculation |
| Range resolution | range_res | Frequency of load measurement for baseline calculation |
| Date range | base_dates | Dates during which load measurements are taken for baseline calculation |
| Function | function | Function applied to load measurement for baseline calculation |
| Firm level demand | firm_level | Load level that companies are expected to reduce their load to (alternative to baseline) |

### S.2.2. Program Data

There are 138 individual programs from 45 states and Washington D.C. States not present include Kansas, Nevada, Wyoming, Maine, and Louisiana. Program data includes entries for each parameter outlined in the metadata for each program. Entries of "n/a" indicate that parameter information is unavailable for that program.

This data can be used to calculate incentive payment for simulated demand response events using a generalizable baseline function and payment function. The baseline function is used to calculate the reference energy usage program participants must reduce their load from to receive incentives. The Payment function will use the baseline energy usage, load reduction, and incentive payment rate to calculate the payment awarded for a demand response event.

### S.2.3. Methods

United States Department of Energy (DOE) data was used to compile the dataset. We reference the US DOE Federal Energy Management Program's Demand Response and Time-Variable Pricing website, which lists and describes energy management programs across the United States to create this dataset (U.S. Department of Energy, 2024a).

First, program parameters were defined. These parameters characterize each program by its distinct features. These parameters can be found in program_parameters.csv in the metadata file.

Next, programs were populated in the dataset along with their parameters (if available) and a URL to the program source or website.

We cross-checked with the US DOE Federal Energy Management Program's Demand Response and Time-Variable Pricing Programs Search after populating the dataset (U.S. Department of Energy, 2024b).

# S.3. Emission and Cost Correlation

## S.3.1. Heat Maps

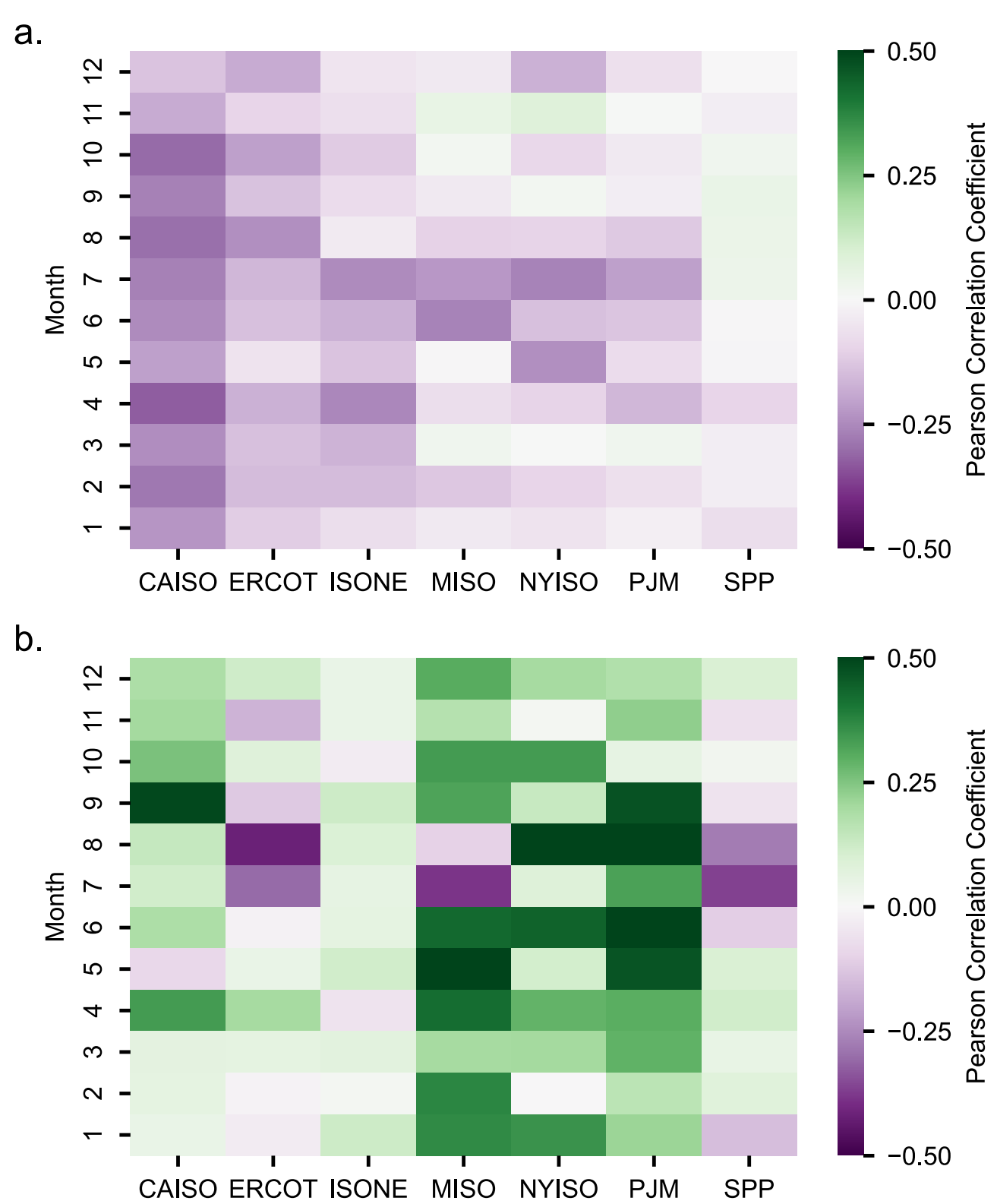

**Figure S3. Heat map of correlation between cost and emissions by month and region across the US in 2023.** a) average emission factor (AEF) and retail electricity tariff correlation. b) marginal emission factor (MEF) and day-ahead market (DAM) price correlation.

Each region has different seasonal behavior with respect to correlation of costs and emissions (Figure S3). For AEF and tariff, a majority of months have negative correlation, while MEF and DAM tend to be more positively correlated. The MEF and DAM heat map seems to more seasonal trends, with a frequent decrease in correlation during the late summer when demand on the grid often peaks.

## S.3.2. Geospatial Visualization

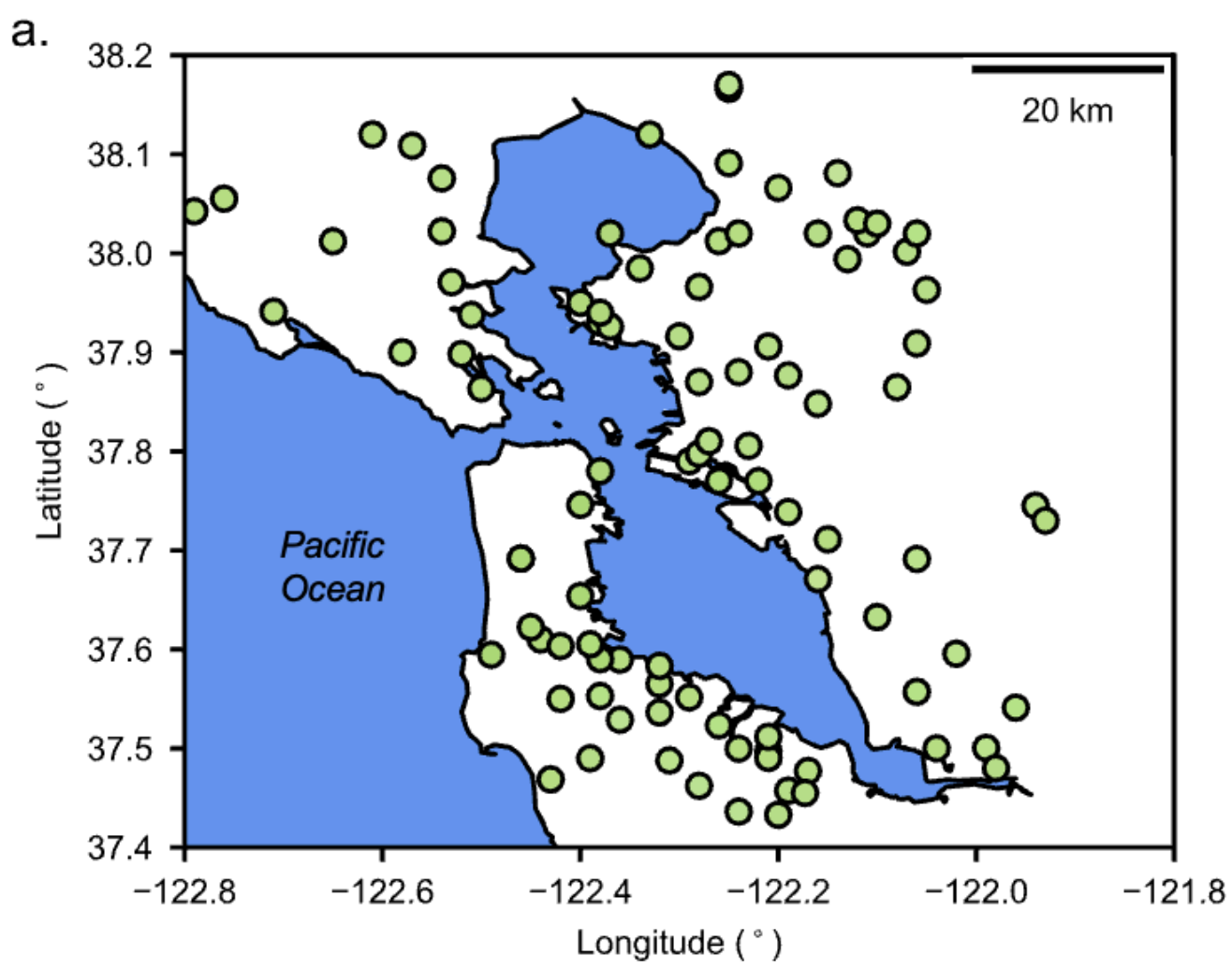

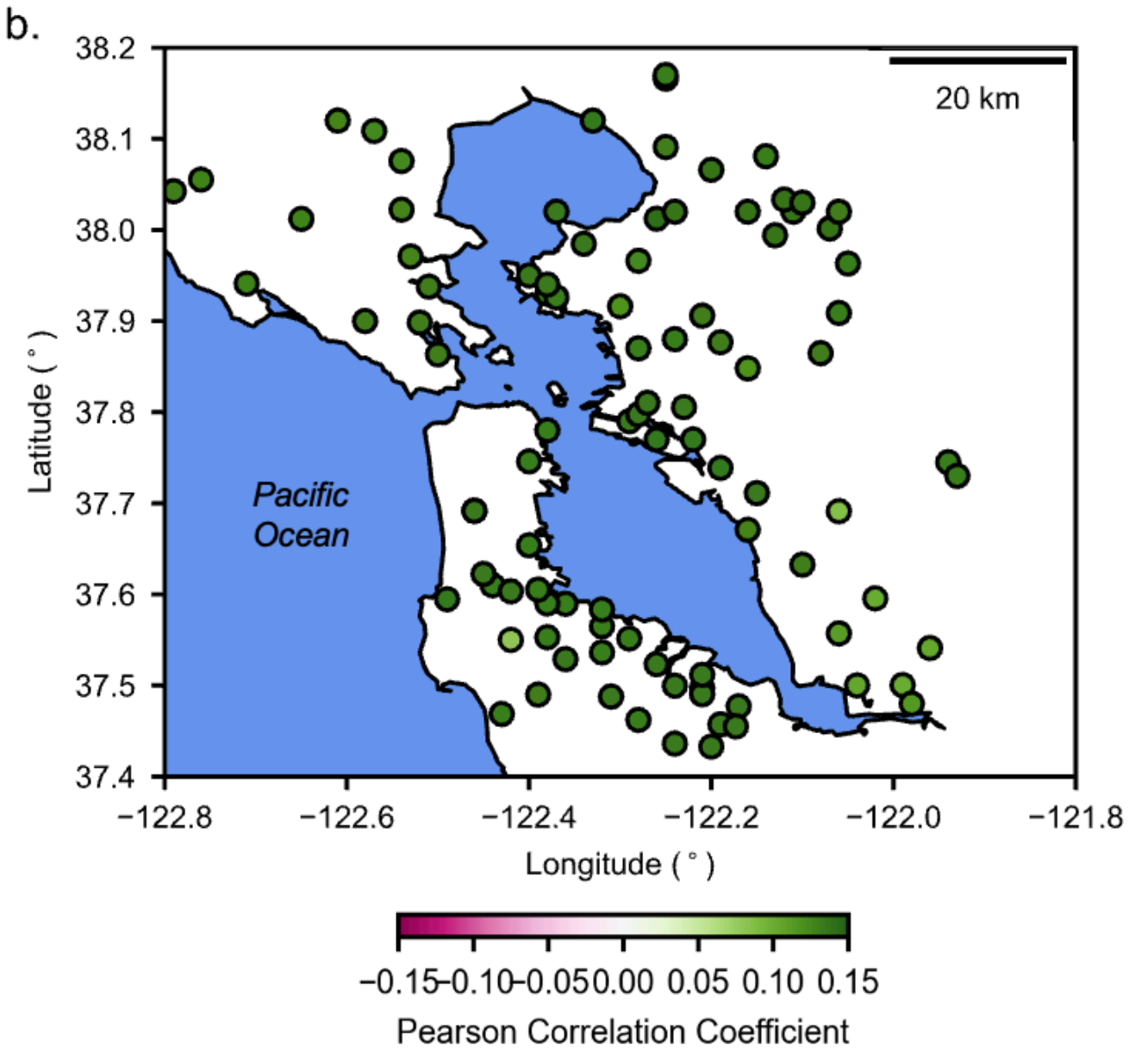

**Figure S4. Nodal variation in correlation between marginal emission factor (MEF) and day-ahead market (DAM) price in the San Francisco Bay Area.** a) Correlation during January 2023. b) Correlation during July 2023. A more positive correlation (close to 1) indicates that price and emissions incentives are aligned, while a negative correlation indicates misalignment.

There is a clear seasonal trend in the correlation between DAM price and MEF in the SF Bay Area (Figure S4). In both winter and summer there is a positive correlation, but in the summer the correlation is larger. The cost/emissions correlation flips (positive to negative or vice versa) for 0% of the nodes in the SF Bay Area between January and July. This contrasts with the 22.5% of nodes that flipped across CAISO. Across the entire United States (i.e., all the nodes shown in Figure 5B), 12.0% of nodes flipped from positive correlation to a negative correlation or from negative correlation to positive correlation between summer and winter. This is likely due to the change in seasonal renewable energy generation and CAISO relatively high percentage of solar and wind generation (Daneshi, 2018).

It should be noted that the geographic spread in correlation coefficients within the SF Bay Area is much smaller in January (0.05 to 0.07) than in July (0.08 to 0.14). This could occur for many reasons, such as distribution congestion, differences in generation type and fuel costs, and non-homogenous power demand (i.e., large loads).

## S.3.4. Seasonal Variation

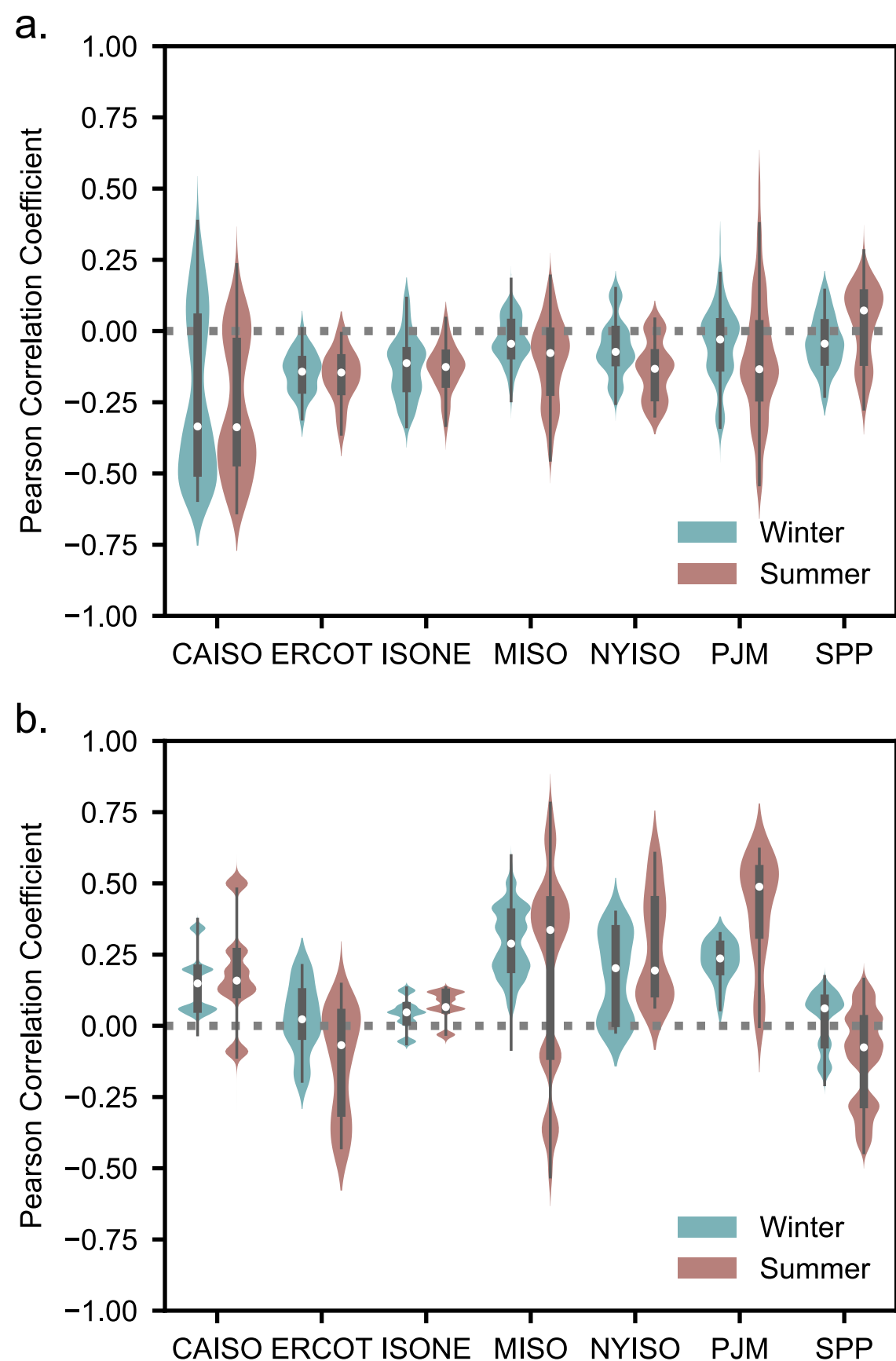

**Figure S5. Correlating prices and emissions in 2023.** a) Pearson correlation coefficient between historical average emission factor (AEF) and time-averaged tariffs, including both energy and demand charges for each ISO and season pair. Each sample is the Pearson correlation coefficient of a different monthly tariff and the average AEF for the region. Flat tariffs are excluded as they have no correlation. b) Pearson correlation coefficients of the average marginal emissions factors (MEFs) for a region and nodal day-ahead market (DAM) prices. Violin plot widths are equal regardless of sample size to highlight the underlying distribution. Winter is defined as November to April, and summer is May to October.

Analysis of seasonal variation is shown in Figure S5. There does not seem to be any seasonal trend for AEF and Tariff, but there are some regions with a seasonal trend in MEF and DAM (e.g., SPP, PJM, and ERCOT) and some without such a trend (e.g., ISONE and NYISO). Month-to-month variation is further elucidated in the heat maps available in Figure S3.

## S.3.4. Alternative Load Shapes

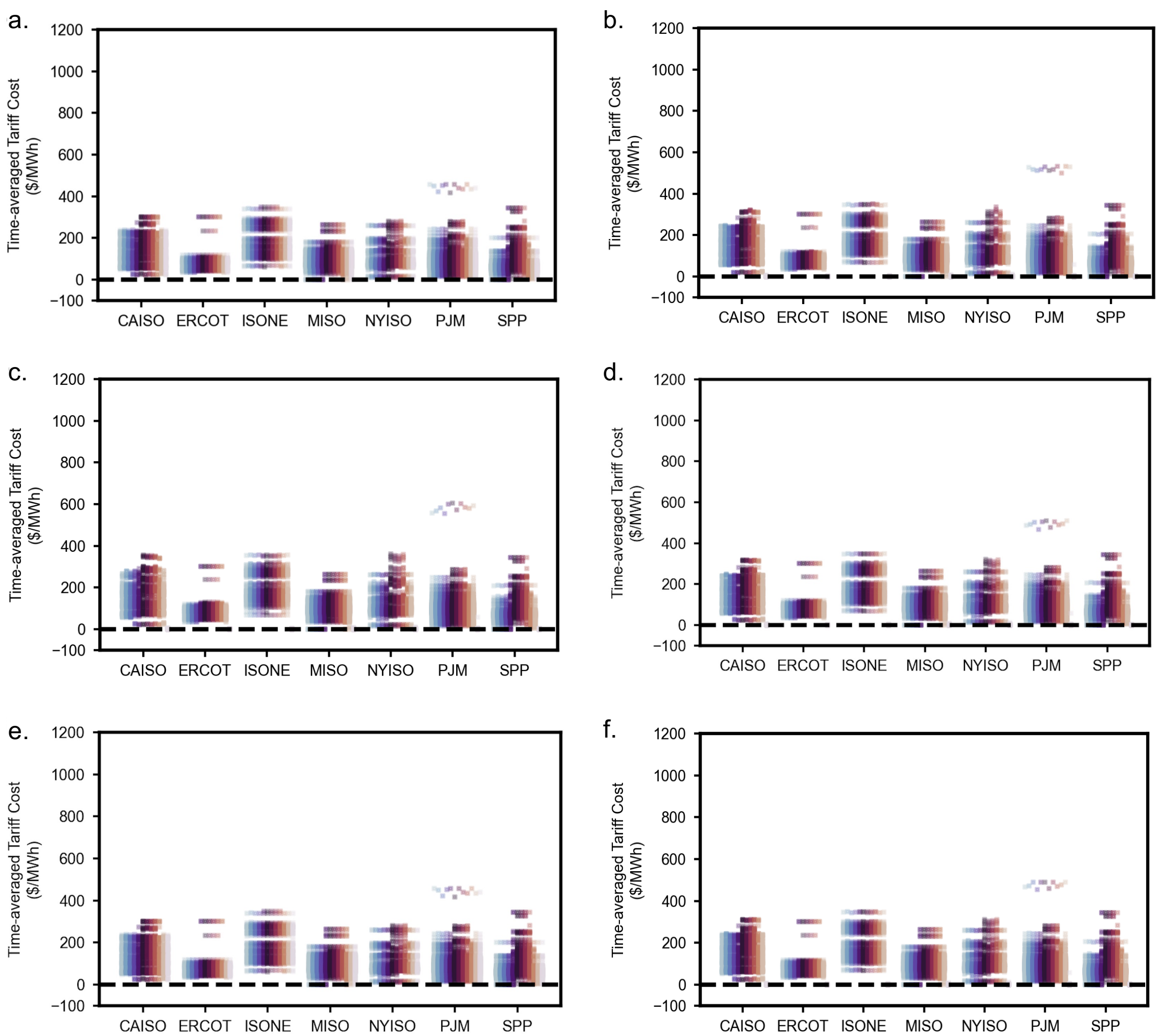

**Figure S6. Time-averaged Tariff Price considers the total average cost of electricity in each ISO/RTO region. The scatter plot shows monthly averages for each region, where the colors of each point represent the month.** a) flat 1 MW load. b) ELMAS cluster 1 (manufacturing) normalized to 1 MW. c) ELMAS cluster 2 (non-food trades) normalized to 1 MW. d) ELMAS cluster 5 (office) normalized to 1 MW. e) ELMAS cluster 6 (crop farming and transportation) normalized to 1 MW. f) ELMAS cluster 8 (water supply and telecommunications) normalized to 1 MW. There are no negative tariffs, but the axis is negative to align with the wholesale price data.

We retrieved load shapes from the ELMAS dataset of French industrial electricity profiles (Bellinguer et al., 2023a). There are 18 clusters of load shapes in the ELMAS dataset. We use the representative load profile for clusters 1 (manufacturing), 2 (non-food trades), 5 (office), 6 (crop farming and transportation), and 8 (water supply and telecommunications) (Bellinguer et al., 2023b). As Figure S6 shows, performing our analysis with different load shapes reveals little difference between the time-averaged cost of electricity for a consumer.

Figure S7 shows a similar lack of variation between the various load shapes. There are slight differences in the distribution of correlation coefficients, but the general patterns within each ISO remain the same.

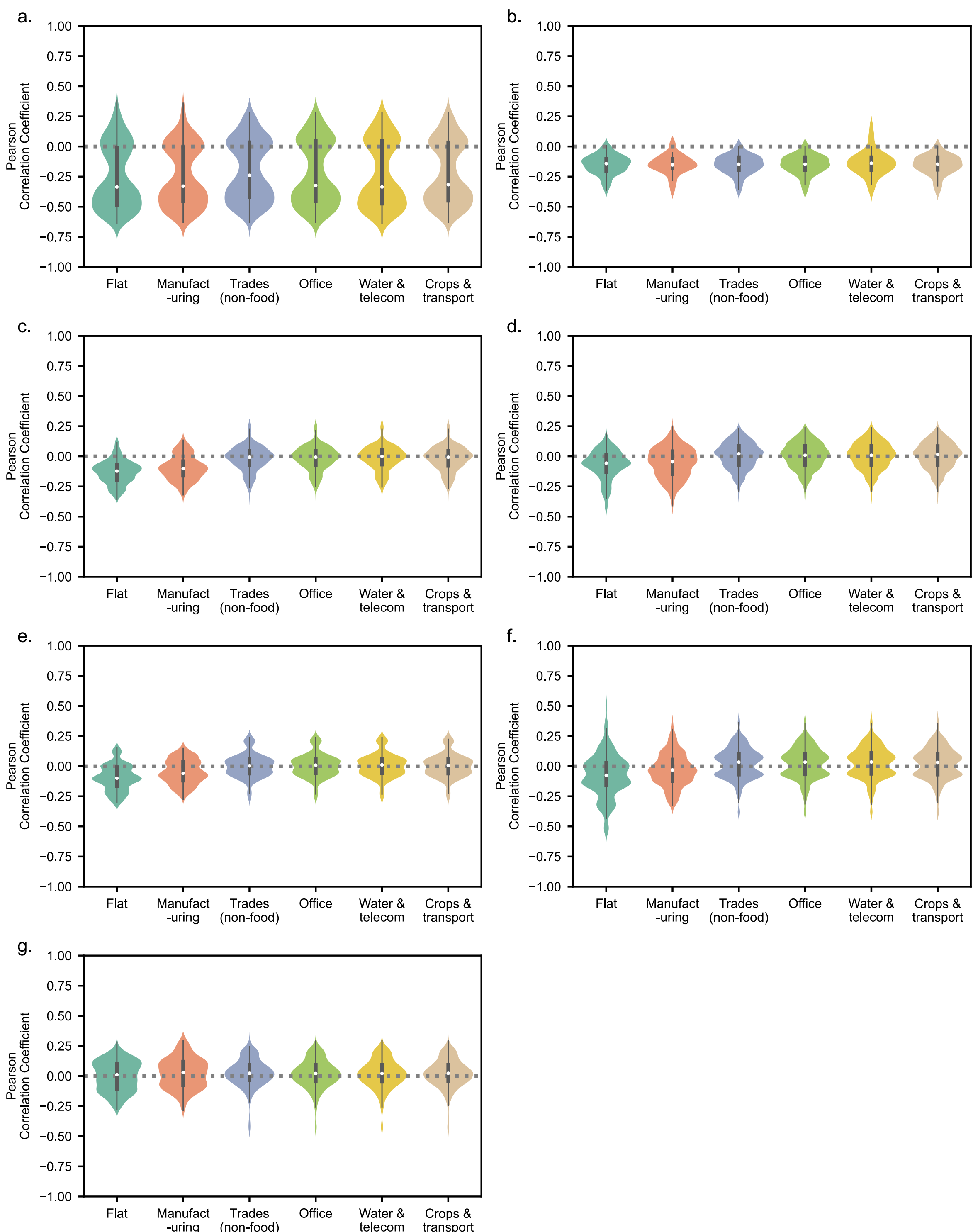

**Figure S7. Violin correlating prices and emissions in 2023 for various load profiles from ELMAS** (Bellinguer et al., 2023a). a) CAISO b) ERCOT c) ISONE d) MISO e) NYISO f) PJM g) SPP. A more positive correlation (close to 1) indicates that price and emissions incentives are aligned, while a negative correlation indicates misalignment.

# References

Bellinguer, K., Girard, R., Bocquet, A., & Chevalier, A. (2023a). ELMAS: A one-year dataset of hourly electrical load profiles from 424 French industrial and tertiary sectors. *Scientific Data*, *10*(1), 686. https://doi.org/10.1038/s41597-023-02542-z

Bellinguer, K., Girard, R., Bocquet, A., & Chevalier, A. (2023b). *ELMAS dataset* (p. 8458390 Bytes) [Dataset]. figshare. https://doi.org/10.6084/M9.FIGSHARE.23889780

Chapin, F. T., Bolorinos, J., & Mauter, M. S. (2024). Electricity and natural gas tariffs at United States wastewater treatment plants. *Scientific Data*, *11*(1), 113. https://doi.org/10.1038/s41597-023-02886-6

Daneshi, H. (2018). Overview of Renewable Energy Portfolio in CAISO - Operational and Market Challenges. *2018 IEEE Power & Energy Society General Meeting (PESGM)*, 1–5. https://doi.org/10.1109/PESGM.2018.8586491

David, E., Sakthivelu, A., Rao, A. K., & Mauter, M. S. (2024). *US incentive based demand response program parameters* (Version 2) [Dataset]. Stanford Digital Repository. https://doi.org/10.25740/ck480bd0124

GridStatus.io. (2025). *Grid Status*. https://www.gridstatus.io/

Midcontinent Independent System Operator. (2023). *Day-Ahead Pricing (xls)*. Market Reports. https://www.misoenergy.org/markets-and-operations/real-time--market-data/market-reports/

Southwest Power Pool. (2023). *Day-Ahead Market*. SPP Portal. https://portal.spp.org/groups/day-ahead-market

U.S. Department of Energy. (2024a). *Demand Response and Time-Variable Pricing Programs*. Federal Energy Management Program. https://www.energy.gov/femp/demand-response-and-time-variable-pricing-programs

U.S. Department of Energy. (2024b). *Demand Response and Time-Variable Pricing Programs Search*. Federal Energy Management Program. https://www.energy.gov/femp/demand-response-and-time-variable-pricing-programs-search